\documentclass[12pt]{article}
\usepackage[utf8]{inputenc}
\usepackage[LGR, T1]{fontenc}
\usepackage{graphicx}
\usepackage{subcaption}
\usepackage{cancel}
\usepackage{float}
\newcommand{\indep}{\perp \!\!\! \perp}
\usepackage{comment}
\usepackage[inline, shortlabels]{enumitem}
\usepackage[reqno]{amsmath} % leqno if want equation number to the left by default
\usepackage{amsthm}
\usepackage{mathtools,bbm}
\usepackage{amssymb}
\usepackage{amsfonts}
\usepackage{eurosym}
\usepackage{geometry}
\usepackage{dsfont}
\usepackage{setspace}
\usepackage{natbib}
\usepackage{sectsty} % section fontsize
\usepackage[colorlinks]{hyperref}
\hypersetup{citecolor=blue,linkcolor=blue, urlcolor=blue}
\usepackage{pgfplots}
\usepackage{tikz}
\usetikzlibrary{patterns}
\usetikzlibrary{datavisualization}
\usetikzlibrary{datavisualization.formats.functions}
\usetikzlibrary{positioning}
\usetikzlibrary{intersections, pgfplots.fillbetween}

\usepackage{subcaption}

\usetikzlibrary{shapes,decorations,arrows,calc,arrows.meta,fit,positioning}
\tikzset{
    -Latex,auto,node distance =1 cm and 1 cm,semithick,
    state/.style ={ellipse, draw, minimum width = 0.7 cm},
    point/.style = {circle, draw, inner sep=0.04cm,fill,node contents={}},
    bidirected/.style={Latex-Latex,dashed},
    el/.style = {inner sep=2pt, align=left, sloped}
}
\usepackage[flushleft]{threeparttable}
\usepackage[multiple]{footmisc} % just for multiple footnote separated with comma

\theoremstyle{plain} % https://en.wikibooks.org/wiki/LaTeX/Theorems
\newtheorem{theorem}{Theorem}%[section]
\newtheorem{assumption}{Assumption}[section]

\newtheorem*{normalization}{Normalization}
\newtheorem{Lemma}{Lemma}

\newcommand*{\myfont}{\fontfamily{qpl}\selectfont}
\DeclareTextFontCommand{\textmyfont}{\myfont}

\makeatletter
\newcommand{\leqnomode}{\tagsleft@true}
\newcommand{\reqnomode}{\tagsleft@false}
\makeatother

\usepackage{xr}
\DeclareSymbolFont{upgreek}{LGR}{cmr}{m}{n}
\SetSymbolFont{upgreek}{bold}{LGR}{cmr}{bx}{n}
\DeclareMathSymbol{\Epsilon}{\mathalpha}{letters}{"0F}
\DeclareMathSymbol{\Eta}{\mathalpha}{letters}{"11}
\DeclareMathSymbol{\epsilon}{\mathalpha}{letters}{`e}
\DeclareMathSymbol{\eta}{\mathalpha}{letters}{`h}

\def\R{{\mathbb{R}}}

\def\P{{\mathbb{P}}}
\def\E{{\mathbb{E}}}

\usepackage{changepage}

\definecolor{ovpurple}{RGB}{122, 33, 130}

\usepackage{titlesec}
\titlespacing*{\section}{0pt}{19pt}{7pt}
\titlespacing*{\subsection}{0pt}{14pt}{5pt}
\titlespacing*{\subsubsection}{0pt}{12pt}{5pt}
\titlespacing*{\paragraph}{0pt}{9pt}{9pt}
\titleformat{\section}{\normalfont\fontsize{16}{15}\bfseries}{\thesection}{1em}{}
\titleformat{\subsection}{\normalfont\fontsize{14}{15}\bfseries}{\thesubsection}{1em}{}
\titleformat{\subsubsection}{\normalfont\fontsize{12}{15}\bfseries}{\thesubsubsection}{1em}{}

\allowdisplaybreaks

\begin{document}

%\newgeometry{top=0.8in} 

\title{{\fontsize{14}{20} \selectfont 
%\vspace{1cm}
\textbf{\textmyfont{%Don't (fully) exclude me, it's not necessary! \\
Identification with possibly invalid IVs}}}}

\author{{\fontsize{12}{20} Christophe Bruneel-Zupanc}\footnote{E-mail address: \href{mailto:christophe.bruneel@gmail.com}{christophe.bruneel@gmail.com}} \\ {\small \textit{Department of Economics, KU Leuven}} \and {\fontsize{12}{20} Jad Beyhum}\footnote{E-mail address: \href{mailto:jad.beyhum@kuleuven.be}{jad.beyhum@kuleuven.be}}\\ {\small \textit{Department of Economics, KU Leuven}} } 
%\vspace{0.25cm} %grantSTG/21/040.} \\

%\vspace{10cm}

\date{{\normalsize \today}\vspace{0.25cm} \\
%{\color{blue}{\normalsize \centering \textbf{Preliminary and Incomplete!}  \textbf{Please do not circulate}}}
%\parbox{\linewidth}{\centering \textbf{Preliminary and Incomplete!}  \textbf{Do not circulate}}
}

%\vspace{-10cm}
\maketitle

%\vspace{-0.20in}

\begin{abstract}

%\onehalfspacing
{\baselineskip=1.1\baselineskip

\noindent This paper proposes a novel identification strategy relying on \textit{quasi-instrumental variables} (quasi-IVs). A quasi-IV is a relevant but possibly invalid IV because it is not exogenous or not excluded. We show that a variety of models with discrete or continuous endogenous treatment which are usually identified with an IV - quantile models with rank invariance, additive models with homogenous treatment effects, and local average treatment effect models - can be identified under the joint relevance of two complementary quasi-IVs instead. To achieve identification, we complement one  \textit{excluded} but possibly endogenous quasi-IV (e.g., "relevant proxies" such as lagged treatment choice) with one \textit{exogenous} (conditional on the excluded quasi-IV) but possibly included quasi-IV (e.g., random assignment or exogenous market shocks). Our approach also holds if any of the two quasi-IVs turns out to be a valid IV. In practice, being able to address endogeneity with complementary quasi-IVs instead of IVs is convenient since there are many applications where quasi-IVs are more readily available. Difference-in-differences is a notable example: time is an exogenous quasi-IV while the group assignment acts as a complementary excluded quasi-IV. \\

}
\end{abstract}

\restoregeometry
\onehalfspacing % check that gives 32 lines/page. Seems ok. 
\setlength{\abovedisplayskip}{6pt} % for skips before/after equations/align
\setlength{\belowdisplayskip}{6pt}

\pagebreak

\section{Introduction} 

A popular quest in empirical economics is to recover the causal effect of a treatment variable $D$ (e.g., education), on an outcome variable $Y$ (e.g., earnings). With nonexperimental data, $D$ and unobserved heterogeneity $U$ (e.g., ability) may not be independent. In this case, causal effects are not characterized by the conditional distribution of $Y$ given $D$, and $D$ is deemed endogenous. A convenient strategy to address endogenous selection is to rely on instrumental variables (IVs). A valid IV must be relevant for the selection and "strongly excluded", i.e., be independent of the unobservables $U$ (exogeneity) and have no direct effect on the outcome (exclusion). However, instruments satisfying these conditions are hard to find in many applications of interest. In particular, the exclusion and exogeneity restrictions are rarely jointly satisfied and are often controversial, even for commonly used instruments. \\
\indent In practice, many variables are not entirely valid IVs, but are nonetheless relevant and satisfy either exogeneity or exclusion but not both: we call these "\textit{quasi-instrumental variables}" (quasi-IVs). 
In this paper, we propose a novel identification strategy relying on two complementary quasi-IVs, $Z$ and $W$, which are invalid IVs on their own but "complementarily valid" taken as a set. The first variable, $Z$, is an \textit{excluded quasi-IV} that is relevant (even when controlling for $U$) and has no direct effect on $Y$, apart from its indirect effect through $D$, $U$, and $W$, but is possibly endogenous with respect to the unobservables, $U$.  The second variable, $W$, is exogenous \textit{conditional on $Z$} and relevant but possibly included. We call it the \textit{exogenous quasi-IV}. 
In addition, complementary validity requires that these two quasi-IVs are jointly relevant for the selection into treatment, i.e., ceteris paribus (conditional on U), the effect of Z on D varies with W (and vice versa). Separate relevance of W and Z is a necessary condition for the joint relevance. If either of the two quasi-IVs turns out to be a valid IV, our identification strategy is still valid. To sum up, we relax the joint exclusion and exogeneity of IVs at the cost of a stronger joint relevance of $W$ and $Z$ for the selection into $D$. 

In many empirical settings, finding two complementary quasi-IVs may be easier than finding one valid IV, since the separate requirements on each of the variables are weaker than exogeneity and exclusion combined. Consequently, the results of this paper open up an array of new identification strategies that may prove useful to practitioners.

 From a theoretical perspective, to demonstrate that complementary quasi-IVs can generally be used as a substitute for IVs by practitioners, we show that the models which are usually nonparametrically point identified with an IV can be nonparametrically point identified with two complementary quasi-IVs instead. In particular, we prove nonparametric identification of quantile models with rank invariance \citep{chernozhukovhansen2005}, additive models with homogenous treatment effects \citep{neweypowell2003,darolles2011nonparametric}, and local average treatment effect (LATE) models \citep{imbensangrist1994, heckmanvytlacil2005}. These models encompass the frameworks of many empirical applications of IVs with discrete or continuous treatment.\footnote{In the quantile model with rank invariance, we allow for general quantile heterogenous effects of $W$. Instead, in the LATE framework, we allow the exogenous quasi-IV $W$ to be possibly included with a homogenous effect (this is similar to the fact that time is typically assumed to have a homogenous effect in the difference-in-differences literature). For the LATE framework, we also need a testable local irrelevance assumption positing that there exists a level of the excluded IV such that the exogenous IV does not affect the propensity score.}
 
 The intuition behind our results is as follows. First, the fact that $W$ is exogenous and $Z$ is excluded implies that the quasi-IVs only have a separable effect on the outcome (separability should here be understood in a general nonlinear sense). Yet, joint relevance means that $W$ and $Z$ also have a "jointly relevant" nonseparable effect on the selection into treatment. Since the effect of the quasi IVs on $D$ is more general than the one they have on $Y$, using general variations of the quasi IVs allows to identify the causal effect of $D$ on $Y$. \\
 \indent This is not the first paper to obtain identification building on some separability arguments (see the literature review for references). However, separability assumptions are generally hard to justify, and as a consequence, less attractive for applied researchers. 
  %Obtaining identification via separability arguments is not new in econometrics (see the literature review for references), but separability assumptions are generally hard to justify, and as a consequence, less attractive for applied researchers.  
 What distinguishes our approach is that the separable effect of the two quasi-IVs on the outcome is not an assumption \textit{per se}, but rather a direct consequence of their complementary validity. Separability is obtained "by construction", from the fact that $W$ and $Z$ violate two different IV assumptions in the general nonparametric models that we study. This is especially appealing for applied researchers, who only need to focus on justifying the validity of the quasi-IVs using economic arguments in order to yield the underlying separability and achieve nonparametric identification of their models. The identification arguments build upon economic narratives, instead of hard-to-justify technical assumptions, similar to how the identification using standard IVs comes from narratives justifying the exogeneity and exclusion restrictions. \\

%\textcolor{red}{The intuition behind our results is as follows. First, the fact that $W$ is exogenous and $Z$ is excluded implies that the quasi-IVs only have a separable effect on the outcome (separability should here be understood in a general nonlinear sense). We stress that this separable effect is not an assumption per se but only a consequence of the complementary validity of the quasi-IVs. Second, the joint relevance of the quasi-IVs implies that they have a nonseparable effect on $D$, which leads to identification through the variation of the quasi-IVs. We note here that using separability to obtain identification is not new in econometrics (see the literature review for references). What distinguishes our approach is that we obtain the required separability from a novel premise that $W$ and $Z$ violate two different IV assumptions.}\\

%In the special case of a linear additive model, the complementarity of the two quasi-IVs is equivalent to positing that the interaction between the two quasi-IVs is a valid IV (see the discussion at the end of Section \ref{sec.add}). Indeed, the interaction is relevant (by joint relevance) but excluded from the outcome equation (by separability of the effect of the two quasi-IVs on the outcome). In a more general nonlinear context, the complementarity provides a general form of exclusion restriction via the separability of the effects of the two quasi-IVs, $W$ and $Z$ on $Y$. Indeed, $Z$ is excluded from $Y$ if we control for the unobservables (and for $D$), while $W$ is excluded from these unobservables by exogeneity.  

\noindent \textbf{Empirical examples of quasi-IVs.}  Let us now provide some general examples of complementary exogenous and excluded quasi-IVs. We expand the discussion below in Section \ref{sec.struc}, where detailed examples are discussed more extensively. Section \ref{sec:illustration} also contains an empirical illustration on the estimation of an Engel curve for food.

Typically, the excluded quasi-IV, $Z$, can be thought of as a "relevant proxy": a proxy for the unobservable, $U$, that is also directly relevant for the treatment, even when controlling for $U$, but whose effect on $Y$ only goes through $U$ and $D$. Many variables fit this description. For instance, past school grades or schooling recommendations by teachers in a returns to schooling example, or more generally, past treatments whose effect on the outcome is superseded by the current treatment in dynamic models with adjustment costs. 

To complement the excluded quasi-IV, one needs to find a complementary exogenous quasi-IV, i.e., a variable which is independent from $U$ conditional on $Z$. Conditional exogeneity differs from exogeneity and could be violated even with unconditionally exogenous $W$, if $Z$ was determined as a function of $W$ for example. 
In practice, a convenient way to satisfy this conditional independence is to find a variable that satisfies the even stronger requirement of being jointly independent from $(U,Z)$ by exploiting the timing of the realization of the variables. 
Typically, unanticipated shocks to local markets or local policy changes should be exogenous with respect to individuals' observed and unobserved characteristics but still affect individuals' outcomes directly. The timing of $Z$ occurring before $W$ guarantees that $Z$ could not depend on $W$ if $W$ was unanticipated. 
Random assignment in randomized experiments are other natural examples of exogenous quasi-IVs  \citep[e.g.][...]{bloom1997benefits, heckman1997making, heckmanetal1998, abadie2002instrumental, schochet2008does}. The randomization guarantees its exogeneity with respect to any pre-treatment variables and that it could not be anticipated beforehand. 
However, these assignments are not necessarily valid IVs because there are often reasonable concerns that they may be included and have a direct effect on the outcome, even after controlling for the treatment. For example, winning a randomly assigned voucher to cover the cost of a private school has a direct effect on the subsequent educational outcomes \citep{angrist2002vouchers}. Indeed, there is an included income effect because the families of winners who attend private schools are richer and the lottery assignment may also directly affect the level of effort exerted by the students. Similarly, the military draft \citep{angrist1990} may have a direct effect on drafted individuals because they may behave differently to avoid going to war (by getting more education) and conscientious objectors may go to prison because they did not comply. 

Importantly, a remarkable example of a setting where our results can be used is (fuzzy) Difference-in-Differences (DiD) (see \citealp{cardkrueger1994,athey2006identification,dechaisemartindhaultfoeuille2018}). In DiD, the group assignment corresponds to our $Z$, it is related to unobserved characteristics but typically excluded from the model given these characteristics. 
Time plays the role of $W$, it is exogenous (conditional on the group assignment) but is allowed to have a direct effect on the outcome.  Moreover, $W$ and $Z$ (time and group) are jointly relevant for the selection into treatment, to the point where, in the sharp design, the interaction $W\times Z$ \textit{is} the treatment. Finally, the identifying assumption in DiD postulates that apart from this joint effect on the treatment, time and group only affect the outcome in a separable manner.\footnote{This is clearly visible in Two-way Fixed Effects (TWFE) regressions, where identification requires not to have interacted time $\times$ group fixed effects.}  The identification approach is the same with quasi-IVs: the only joint effect of $W$ and $Z$ on the outcome is indirect through their joint effect on the treatment. 
Therefore, our results can also be seen as a generalization of DiD, extending its applicability to general variables (the quasi-IVs) in place of time and group assignment. Moreover, our identification results also contribute to the literature on identification in DiD models since some of them are new in the context where $W$ and $Z$  are the time and group assignment. We elaborate on this contribution through remarks in the text.

\indent On a final note, remark that our results provide a new identification strategy for policy evaluation. 
Indeed, policy changes can serve as valid exogenous quasi-IV. Our results therefore also allow us to identify the direct effect of a policy change on the outcome, net of its indirect effect through the selection into the treatment. To do so, one needs to find an appropriate excluded quasi-IV to complement the policy (e.g., pre-determined individuals' characteristics). \\

\noindent \textbf{Related literature.}  This paper is related to a strand of the econometric literature identifying econometric models with endogeneity without (explicitly) relying on instrumental variables. These papers address endogeneity by imposing (semi)parametric functional form restrictions \citep[see][]{rigobon2003identification,dong2010endogenous,klein2010estimating,lewbel2012using,escanciano2016identification, benmoshe2017identification, lewbel2018identification,bun2019ols,jiang2022consistent,tsyawo2023feasible,gao2023iv,lewbel2024identification}. These restrictions serve to confine the structural regression function's degrees of freedom, enabling the identification of the causal effect of the endogenous variable through nonlinear variations induced by exogenous (yet included) variables. Our approach differs from this literature on two dimensions. First, we obtain nonparametric identification, while the aforementioned papers typically rely on semiparametric models. Second, we do not assume separability but rather show how it results from the properties of the quasi-IVs. Justifying the validity of the quasi-IVs using economic arguments (similarly to how researchers argue for the exclusion and exogeneity of standard IVs) allows to indirectly justify the separability. \\
%justify it economically through the properties of the quasi-IVs.\\
%
\indent We note that separability-type restrictions have also been used in settings with standard (excluded and exogenous) instruments to obtain identification when the relevance condition may not hold (including the case where there are insufficiently many IVs), see \cite{caetano2021identifying, huang2019identification} and \cite{feng2024matching}, among others.\footnote{\cite{caetano2021identifying} treats a different problem than ours but some of the tools developed in \cite{caetano2021identifying} could be used in our context. In particular, in the quantile model of Section \ref{sec.ivqr}, we use a conditional independence assumption, which could be replaced by a conditional zero covariance assumption as in \cite{caetano2021identifying}. While conditional zero covariance is weaker than conditional independence, we would then need a conditional completeness relevance assumption as in \cite{caetano2021identifying} for identification. This is stronger than the relevance condition we impose in Section \ref{sec.ivqr} since our condition can use more moments to obtain identification. Moreover, in the case where $Z$ and $W$ are discrete, our relevance condition is easily understandable and interpretable.} Here, we rather use separability to address a lack of exogeneity or exclusion. From a general perspective, separability is a common tool in econometrics. For instance, the separability of fixed effects with respect to time and subjects allows the identification of panel data models with two-way fixed effects (including DiD models). \\
\indent Another line of works studies the identification of causal effects using imperfect (endogenous and/or included) instrumental variables satisfying different assumptions than our quasi-IVs. \cite{kolesar2015identification} considers the case where there are many included IVs whose direct effects on the outcome are uncorrelated with their effects on the treatment. \cite{liu2020two} point identifies the partial derivative of the average treatment effect with respect to an exogenous quasi-IV, but they need this included instrument to be excluded from the selection equation. 
\cite{DHAULTFOEUILLE2024105075} adopts a control function framework using an exogenous quasi-IV, achieving point identification through a local irrelevance condition. We employ a similar restriction to identify our LATE model. \cite{wang2023point} point identifies the LATE with an exogenous quasi-IV satisfying the standard monotonicity assumption of \cite{imbensangrist1994} and another exogenous and excluded IV that can violate monotonicity. 

Notice also that our excluded variable $Z$ can be thought of as a relevant proxy. Proxy variables have a long history in econometrics and statistics and are the subject of renewed attention, see \cite{deaner2018proxy}, \cite{ying2023proximal} and references therein. This literature considers identification with several proxy variables while we rely on a single proxy $Z$, that needs to be relevant, complemented by the variable $W$ which is not a proxy. \\  %In the case of discrete treatment, the two approaches are complementary, and one could use one semi-IV in place of the included IV since a semi-IV is also exogenous for example. 
% MAYBE ADD A SENTENCE HERE TO RECLARIFY THE DIFFERENCE FOR US? 
\indent The present paper is also related to \cite{bruneel2023don}, which introduces semi-instrumental variables (semi-IVs) in the context of discrete treatment variables. A semi-IV is an exogenous variable that is only excluded from some (but not all) potential outcomes. Identification with semi-IVs works for similar reasons as identification with complementary quasi-IVs: it requires complementary semi-IVs, i.e., at least one semi-IV excluded from each potential outcome (hence a complementarity in the exclusion), satisfying a stronger joint relevance (for the quantile model with rank invariance) or larger support condition (for the LATE) than standard IVs. The two papers are related but study identification with a completely different type of invalid IVs. The advantage here is that quasi-IVs also apply to models with continuous endogenous variables while semi-IVs cannot by construction (because one would need to find infinitely many of semi-IVs for infinitely many potential outcomes). The assumptions, results and proofs of the two papers differ.  \\
\indent Finally, while we focus on point identification with two complementary quasi-IVs, some other works bound the treatment effects with an invalid IV, see \cite{manski2000monotone,nevo2012identification,conleyetal2012,flores2013partial,mealli2013using,chesher2020generalized,ban2022nonparametric} among others. \\
\indent Concerning the terminology, \cite{bartels1991} also uses the term "quasi-IV", but referring to variables that can be both included and endogenous, but only slightly deviating from the exclusion/exogeneity requirements. Then, he conducts a sensitivity analysis in this setting. In our case, the prefix "quasi" means that the variables only satisfy part of the IV requirements, but in the dimension from which they deviate from the standard, the violations of the exclusion or exogeneity can be very large. \\

 \noindent \textbf{Outline.} The paper is organized as follows. In Section \ref{sec.ivqr}, we show identification in a quantile model with rank invariance and discuss further concrete examples of quasi-IVs. Then, Section \ref{sec.add} contains identification results for an additive model with homogenous treatment effects. The LATE framework is discussed in Section \ref{sec.late}.  Section \ref{sec:illustration} presents an empirical illustration on the estimation of an Engel curve for food. Section \ref{sec.ccl} concludes. The proofs of our formal results, along with an extension of the results of Section \ref{sec.late} to generalized LATE and marginal treatment effects parameters can be found in the Online Appendix.

\section{Quantile model with rank invariance}\label{sec.ivqr}
\subsection{The model}\label{section_framework}
In this section, we consider identification with quasi-IVs in a quantile model with rank invariance similar to the instrumental variable quantile regression (IVQR) model of \cite{chernozhukovhansen2005}. Let $D $, with support $\mathcal{D}$,
be the observed (endogenous) choice/treatment. Let $Z$, with support $\mathcal{Z}$, be an excluded (possibly endogenous) quasi-IV, whose possible effect on the outcome only goes through $D,W$ and the unobservables $U$. Denote by $W$, with support $\mathcal{W}$, the complementary exogenous quasi-IV (conditional on $Z$), which may be "included" and have a direct effect on the outcome. 
Let $Y_{dw}$ the (continuous) latent potential outcome under treatment state $d\in\mathcal{D}$ and value of the exogenous quasi-IV equal to $w\in\mathcal{W}$. The potential outcomes $\{Y_{dw}\}$ are latent because we only observe one outcome, $Y=Y_{DW}$, corresponding to the potential outcome of the selected alternative $D$ and the realized value of $W$.
By construction, the effects of $W$ and $Z$ on $Y$ are separable, in the sense that, if we could control for $U$ (in addition to $D$ and $W$), $Z$ would have no effect on $Y$. 

The framework is implicitly conditional on additional covariates $X$, which are generally neither excluded nor exogenous.  All our results and assumptions can be interpreted as conditional on $X$. We do not explicitly condition on $X$ to simplify the exposition. 

We study the following structural quantile regression model with endogeneity:
\begin{align}
\label{baseline1} Y_{dw}&=h(d,w,U),\ U \sim\mathcal{U}[0,1],
\end{align}
for all $d\in\mathcal{D}$, where $h$ is strictly increasing in its last argument. The treatment is possibly endogenous since we do not assume that $U\indep D|W$. $Z$ is excluded from the potential outcomes given $W$ and $U$.

To address the endogeneity issue, we assume that there exists a mapping $g: \mathcal{Z}\times[0,1]\to \R$ increasing in its second argument and a random variable $V$ such that 
\begin{align}
\label{baseline2} U&=g(Z,V),\ V|Z,W\sim\mathcal{U}[0,1].
\end{align}
Equation \eqref{baseline2} is equivalent to our key identification condition $$U\indep W| Z,$$ that is $W$ is exogenous given $Z$.\footnote{Indeed, when \eqref{baseline2} holds, conditional on $Z$, the distribution of $U$ is fully determined by that of $V$ which is independent of $W$, so that $U\indep W| Z$. When we have $U\indep W| Z$, we obtain \eqref{baseline2} by taking $g(z,v)=\inf\{u\in[0,1]:\ F_{U|Z}(u|z)\ge v\}$ and $V=F_{U|Z}(U|Z)$, where $F_{U|Z}(u|z)=\P(U\le u|Z=z)$.} However, $W$ is not excluded from \eqref{baseline1} and $Z$ can be endogenous through \eqref{baseline2}. If they also satisfy a joint relevance condition, $W$ and $Z$ are "complementarily valid" quasi-IVs in this framework. 
The facts that $U$ and $V$ have a uniform distribution are just some normalizations.
 Note that, for $d\in\mathcal{D},w\in\mathcal{W},z\in\mathcal{Z}$ and $v\in[0,1]$,  $h(d,w,g(z,v))$ is the $v$-quantile of the distribution of $Y_{dw}$ given $Z=z$. The model in \eqref{baseline1} implies rank invariance \citep[see][]{chernozhukovhansen2005}. It essentially means that the rank in the outcome of any two subjects is the same across all potential outcomes.\footnote{We could relax this assumption and instead impose rank similarity as in \cite{chernozhukovhansen2005}, but we do not do so to avoid complicating the exposition.} \\
\indent Importantly, notice that we allow $W$ to be excluded and/or $Z$ to be exogenous. It nests the case where any (or both) of the two quasi-IVs turns out to be a valid IV. 

\begin{figure}[!h]
\centering
\caption{Causal graph of the Framework}\label{fig_graph}
%\captionsetup[subfigure]{font=footnotesize}

\centering
\begin{tikzpicture}[node distance={25mm}, thick] 
% Nodes:
\node (D) at (0, 0) [label=below:$D$, point]; 
\node (Z) at (-1.5, 1) [label=above left:$Z$, point];
\node (W) at (-1.5, -1) [label=below left:$W$, point];
\node (Y) at (1.5, 0) [label=right:$Y$, point]; 
\node (h) at (0, 2) [fill=white, label=above:$ U$, point];

\path (D) edge (Y); 
\path (Z) edge (D);
\path (W) edge (D);
\path (W)[dotted] edge[bend right=20] (Y);
%\path (Z)[-, dotted] edge[bend right=20] (W);
%\path (W)[dashed] edge[bend right=20] (Z);
\path (h)[-, dotted] edge (Z);
\path (W)[-, dotted] edge (Z);
\path (h)[dotted] edge (D);
\path (h)[dotted] edge (Y);

%\draw[dashed, ->] (W) to [out=0,in=0,looseness=0] (Y); 
\end{tikzpicture} \\ %\\

\vspace{10pt} 

{\scriptsize{}}%
\noindent\begin{minipage}[t]{1\columnwidth}%
{\footnotesize{} \textit{Solid and empty nodes represent observed and unobserved variables, respectively. Solid arrows indicate causal effects. The dotted arrows indicate possible causal relationships. The dotted undirected lines indicate possible general relationships without specifying the direction.}}%
\end{minipage}{\footnotesize\par}
\end{figure}
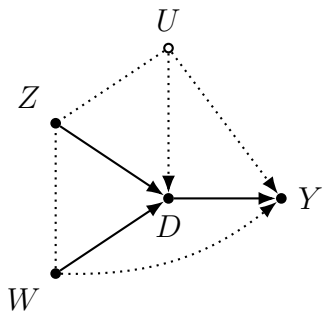

Our framework can be visualized through the causal graph of Figure \ref{fig_graph}.
The fact that the only link between $W$ and $U$ passes by $Z$ means that $W$ and $U$ are independent given $Z$, ensuring conditional exogeneity of $W$. There is no direct arrow from $Z$ to $Y$, so that $Z$ is excluded given $D,W$ and $U$. $W$ and $Z$ can be correlated, even though, in most applications, they will not be. We caution the reader that causal graphs are only a visualization tool but do not rigorously encode all probabilistic assumptions.

Notice that model \eqref{baseline1}-\eqref{baseline2} can be thought of as a structural model where $U$ is a structural error  (e.g., unobserved ability, preference, productivity, ...). Alternatively, one could write a quantile framework with rank invariance for quasi-IVs starting directly from the potential outcomes. We do so in Section \ref{sec.alternative} of the Online Appendix. The two frameworks are equivalent when it comes to the identification arguments.

\subsection{Discussion and structured examples}\label{sec.struc}
Let us discuss the conditions under which $W$ and $Z$ are complementarily valid quasi-IVs and provide several examples of applications. Every statement is implicitly conditional on $X$, and we only mention $X$ when necessary.

%\noindent \textbf{The difference between $\bf{W \indep U}$ and $\bf{W \indep U | Z}$.} \\
Our framework requires conditional independence, that is $W \indep U | Z$. This is different from unconditional exogeneity, $W \indep U$. Finding an exogenous $W$ is not sufficient (nor necessary), one needs to find a $W$ that is exogenous when conditioning on $Z$. In particular, we may run into a problem if $Z$ is a function of $W$, i.e., $Z = \varphi(W, U)$. In this case, even if $W$ is unconditionally exogenous, i.e., $W \indep U$, we will not generally have $W \indep U | Z$. In practice, thinking in terms of conditional independence is often nontrivial. Therefore, to find valid $W$ and $Z$, we suggest looking for variables satisfying the stronger but more intuitive requirement that $W \indep (U, Z)$.\footnote{Nonetheless, it is also possible that $W \indep U | Z$ even if $W \not\indep U$. A typical example of this is when $W$ represents a randomized treatment, but the randomization occurs on the basis of a specific variable, e.g., unemployment status, education, or income. Then, taking these variables as $Z$, we have $W \indep U | Z$, and the requirement is then that the $Z$ are plausibly excluded conditional on $D, U$ and $W$. Therefore, our work implies that randomized experiments can be "conditionally randomized" on the basis of some observable excluded quasi-IVs to favor some target population (e.g., the unemployed or uneducated) and still identify treatment effects for everyone. If one expects positive effects from the treatments, being able to identify the treatment effect with this conditional randomization is a desirable property, especially when the program is costly.}

To ensure this stronger condition when $Z$ is a choice, we need $W$ to be excluded from the decision $Z$. This can be achieved by exploiting the timing of the variables, using a $W$ which is unknown and unanticipated when $Z$ is decided. For example, in a dynamic model where $Z$ is a choice taken in period $t-1$, our assumption holds if $W$ is an exogenous shock occurring in period $t$ that was unanticipated in period $t-1$. This is why, in many examples below, we use "past decisions/outcomes" as $Z$, and exogenous shocks (i.e. contemporaneous variations) to the local market conditions as $W$. We use unexpected shocks to the local market conditions rather than these local market conditions themselves as $W$ because $Z$ could have been chosen based on past local market conditions which  correlate with current local market conditions. 
\indent Now, we provide 6 examples of applications of our approach. They illustrate the type of empirical patterns that can be studied through our novel identification results. Another example on the estimation of Engel curves can be found in our empirical illustration presented in Section \ref{sec:illustration}.\\

%\noindent \textbf{Examples:} \\
\noindent \textbf{Example 1 (Returns to schooling).} %\\ 
We are interested in the effect of schooling (years of education, or a binary decision to go to college or not), $D$, on the earnings at age $30$, $Y$ \citep{card1995, card2001}. We have an endogeneity issue because the decision to go to college and the subsequent earnings are both impacted by unobserved ability, $U$. To identify the effect of $D$ on $Y$, we need to find two complementary quasi-IVs. Let us consider past grades or school recommendations by teachers as the excluded quasi-IV, $Z$. These variables are "relevant proxies" in the sense that they are likely strongly correlated with individuals' unobserved abilities ($U$), but once you control for $U$ and $D$ (and $X$), they have no effect on the outcome. Moreover, they are still relevant, even controlling for $U$, because, everything else equal (i.e., at fixed ability), a student with better grades or with a better recommendation is more likely to go to college. We complement these excluded IVs by using exogenous shocks to the local market conditions (wage levels, unemployment rates, tuition fees, ...) at the time of the decision to go to college (around age $17$) as the exogenous quasi-IV, $W$. These types of exogenous shocks were used as IVs by \cite{carneiroheckmanvytlacil2011} when controlling for permanent market conditions, for example. However, these $W$ may not be valid IVs because they may be included if, for example, a negative shock to local market condition at age $17$ persists over time and leave long-lasting scars on the local economy. Yet, they are valid $W$ because they are likely exogenous with respect to unobserved student ability, $U$, and past individuals' grades/school recommendations, $Z$, so $W \indep (U, Z)$, which is stronger than the required $W \indep U | Z$. Notice that it is essential that $W$ was unanticipated at the time when the student grades were obtained. Otherwise, some students may have potentially worked harder because of $W$, and the conditional independence of $W \indep U | Z$ would not necessarily hold then.
The two complementary quasi-IVs are also likely to be jointly relevant if the likelihood of pursuing education for students with good grades compared to students with bad grades varies with market conditions. \\

\noindent \textbf{Example 2 (Returns to private schooling using randomized lotteries).} 
\noindent Another typical question is the returns to private vs. public schooling ($D$) on educational attainment, $Y$. Again, the main problem is that both the educational attainment and the decision to go to a private school are endogenous with respect to the unobserved student ability, $U$. The Colombian government ran a large-scale randomized experiment, the PACES program, giving vouchers that covered a large fraction of the cost of private secondary school \citep{angrist2002vouchers}. The vouchers were assigned randomly via a lottery, so their assignment is exogenous by construction, and it is a natural candidate exogenous quasi-IV, $W$. However, despite its exogeneity, $W$ is probably not a valid IV if we want to study the returns to private schooling because the vouchers may be included and have a direct effect on educational attainment. For example, conditional on going to private education, families of students receiving the vouchers are more affluent than families who did not receive it. This income effect may directly affect the subsequent educational attainment of the children. Another concern is that students with vouchers may exert less effort than those without vouchers to get into the same schools. Because of these concerns, \cite{angrist2002vouchers} study the effect of winning the lottery on educational attainment, not the effect of private schooling itself. Fortunately, using our methodology, we can leverage the exogeneity of $W$ to identify the causal effect of private schooling. To do so, we need to find an excluded quasi-IV, $Z$, to complement the exogenous voucher assignment, $W$. For the same reasons as in the previous returns to schooling example, variables such as pre-lottery student grades are valid $Z$: they are relevant for the selection into private school, even controlling for unobserved ability $U$, but are otherwise excluded from the subsequent schooling attainments when one controls for $D$ and $U$. Moreover, because the lottery was randomized, and especially if the PACES program was unanticipated, $Z$ is also independent of the voucher assignment, $W$. So we have $W \indep (U, Z)$, and thus, $W \indep U | Z$. The joint relevance is likely to be satisfied, as the increase in the probability of attending private school when receiving a voucher should vary with the past school achievements, $Z$. In general, being able to use randomly assigned $W$ that affect the outcome simplifies the task of designing randomized experiments.  \\

\noindent \textbf{Example 3 (Marginal returns to hours worked).} %\\
Suppose we want to identify the marginal effect of working more on earnings, $Y$. In this case, $D$ represents the (continuous) worked hours or a general measure of work flexibility. The main endogeneity issue is that the hours worked decision depends on individuals' unobserved productivity, $U$. To address this endogeneity concern, we need two complementary quasi-IVs.
If we focus on a sample of married men, let us use the spouse characteristics (e.g., spouse education or income) as $Z$ \citep[e.g.,][]{mroz1987}.
$Z$ is correlated with the wife's unobserved productivity, and in the presence of assortative matching, the wife's unobserved productivity may correlate with the husband's unobserved productivity, $U$. It is also possible that the wife's characteristics have an effect on $U$ directly. As a consequence, $Z$ may not be a valid IV because it may not be exogenous. However, $Z$ should be a valid excluded quasi-IV for the following reasons. First, it is likely very relevant for the husband's working hours decision, even controlling for his unobserved productivity, because it has a direct effect on the household available income. Second, controlling for the unobserved productivity, $U$, for hours worked, $D$, and for other control variables such as the husband's experience, $X$, and the wife's characteristics, $Z$, should have no direct effect on the husband earnings, $Y$. $Z$ is thus a plausible "relevant proxy" and therefore excluded quasi-IV.
To complement this excluded quasi-IV, we can use exogenous shocks to the local market characteristics as $W$. The conditional independence condition, $W \indep U | Z$, is more likely to hold if the decision to marry was independent of $W$. It is even more plausible if the marriage happened in a previous period while $W$ are contemporaneous market shocks. As for the joint relevance, it is likely to hold as the marginal effect of $Z$ on $D$ should vary with the market shocks $W$. \\

\noindent \textbf{Example 4 (Dynamic models with adjustment costs).} 
Let us consider dynamic models with adjustment costs.
For example, one can think of a model of labor supply, $D_t$, and consumption, $Y_t$ to study the marginal propensity to consume with respect to work \citep{bcms2016, blundellpistaferisaporta2016, bruneel2022}.\footnote{One could also consider the identification of dynamic production functions (\citealp{olleypakes1996, blundellbond2000, levinsohn2003estimating, acf2015, gandhi2020identification}, see \citealp{deloeckersyverson2021} for a recent survey) with adjustment costs on the inputs as a similar alternative example.} There is an endogeneity issue because the unobserved individual preference for consumption, $U_t$, affects both decisions $D_t$ and $Y_t$.
In this setting, $D_{t-1}$ is a natural excluded quasi-IV, $Z_t$. Indeed, conditional on the current $D_t$, and controlling for current covariates such as wealth and hourly wage in $X_t$, the past labor decision, $D_{t-1}$ does not affect the current consumption choice, $Y_t$. Moreover, in the presence of adjustment costs in the labor supply, the current labor supply decision is directly affected by the previous labor supply choice, even controlling for $U_t$ and $X_t$. $D_{t-1}$ is, however, not a valid IV because it is likely correlated with $U_t$ as soon as there is some persistence in the preference for consumption.
We complement this excluded quasi-IV with contemporaneous exogenous shocks to the local market conditions (e.g., to prices), which were unpredictable at time $t-1$ and are independent of the individual unobserved preferences. Thus, $W_t \indep (U_t, Z_t)$. As stressed earlier, it is important that $W_t$ was unanticipated at $t-1$ when $D_{t-1}$ was decided because otherwise $D_{t-1}$ would depend on $W_t$.
Finally, joint relevance should be satisfied if the effect of $D_{t-1}$ on $D_t$ varies with the contemporaneous exogenous shocks, which is likely to hold: finding a job during a recession is harder.\\

\noindent \textbf{Example 5 (Job Training Programs).} 
We are interested in the effect of job training programs such as JTPA \citep{bloom1997benefits, heckman1997making, heckmanetal1998, abadie2002instrumental} or Job Corps \citep{schochet2008does} on subsequent job market outcomes, $Y$ (employment or earnings). In this case, random assignment into the program is exogenous by construction. It is a valid IV, but only if we want to recover the global effect of the program. However, job training programs typically offer a variety of trainings. Suppose, for example, that everyone receives job search assistance, but only some individuals choose to follow a schooling training in addition. If we want to isolate the effect of the schooling training, $D$, on $Y$, the random assignment to the program is not a valid IV anymore. Indeed, despite its exogeneity, it is included since assigned individuals also obtain job search assistance, which may have a direct effect on the outcome. A solution to address this issue is to find a complementary excluded quasi-IV, $Z$. A good candidate in this setup is pre-assignment schooling achievement. If we consider $D$ as the schooling level after training, $Z$ is excluded from $Y$ conditional on $D$ and $U$ because only the final schooling level matters on the labor market. 
%the subsequent schooling level supersedes the previous one. 
Previous schooling achievement is, however, relevant for the selection into the training since everything else equal, individuals with different schooling backgrounds may be more or less likely to accept participation in the training. It is, however, not a valid IV since it is strongly correlated with unobserved ability, $U$. Since $Z$ is a pre-treatment variable and the random assignment was not anticipated, $W \indep (U, Z)$. Moreover, the joint relevance is satisfied since the increased probability of following the treatment if assigned to it should vary with the previous schooling achievement. So, $W$ and $Z$ are complementarily valid quasi-IVs.\footnote{Treatment recommendations or measures of long-term unemployment probability by the training agencies are also typically good $Z$. One concern here is that when there are several kinds of treatments, the recommendations may not be excluded from $Y$. Indeed, an individual to whom we recommended to follow a school training may follow another non-mandatory training if assigned to the program. As a consequence, assigned individuals ($W=1$) with a schooling recommendation ($Z=1$) and assigned individuals with another recommendation ($Z=0$) do not follow the same training, and the interaction between $W$ and $Z$ has a direct included effect on the outcome. This is not a concern if the alternative training is mandatory and followed by everyone. Otherwise, we need to control for the different kinds of training (if observable). }  \\

%\newpage
\noindent \textbf{Example 6 (Difference-in-differences).}  Consider a DiD setting where $W$ is time and $Z $ is the group assignment.
$D$ is typically a binary treatment implemented after a given date and only in some treated groups. Our key assumption $U\indep W|Z$ is standard in the DiD literature. Indeed, \cite{athey2006identification} imposes "time invariance within groups" (their Assumption 3.3) and \cite{dechaisemartindhaultfoeuille2018} assumes "time invariance of unobservables" (see their Assumption 7). It states that time is independent of unobserved heterogeneity within each group or, in other words, that the composition of the groups does not change with time. This is exactly $U\indep W|Z$ in our model. In DiD, this assumption is, for instance, satisfied by construction if one observes all units in each group at every date. %\\
\noindent Compared to \cite{athey2006identification} and \cite{dechaisemartindhaultfoeuille2018}, we impose rank invariance with respect to both time and treatment. Instead, they only impose rank invariance with respect to time but not with respect to treatment (see their Assumptions 3.1 and 7, respectively). However, in fuzzy designs where units in a given group can have different treatment statuses, the approach of these two papers can only identify local quantile treatment effects. The results in the present section allows us to identify quantile treatment effects over the whole population instead. We also note that in DiD papers, treatment, time, and group assignments are usually all discrete while we allow them to be continuous here. A notable exception is \cite{d2023nonparametric}, which allows for continuous treatment imposing rank invariance only with respect to time. Unlike us, they need a crossing points assumption to identify the model. Instead, we rely on rank invariance with respect to both the treatment and time. \\
\indent Note that we continue this analysis of our results in the context of DiD in the LATE framework in Section \ref{sec.late}. \\

\subsection{Identification of the model}\label{sec_ID}
\subsubsection{System of equations}
In this section, we study the nonparametric identification of different unknown (infinite-dimensional) parameters. For $u\in[0,1],z \in\mathcal{Z},w\in\mathcal{W}$, let $F_{U|Z}(u|z)=\P(U\le u|Z=z)$ be the conditional distribution of $U$ given $Z$. We want to identify the functions $h$ and $g$.
Notice that, since $V|Z\sim\mathcal{U}[0,1]$, for all $v\in[0,1]$ and $z \in\mathcal{Z}$, we have $g(z,v)=\inf\{u\in[0,1]:\ F_{U|Z}(u|z)\ge v\}$, i.e., $g(z,\cdot)$ is the generalized inverse of $F_{U|Z}(\cdot|z)$. Because of this one-to-one relationship between $g$ and $F_{U|Z}$, we focus on the identification of $F_{U|Z}$ rather than of $g$, but the two are equivalent. It means that we nonparametrically point identify the entire model. This includes the effect of the treatment and of the exogenous quasi-IV $W$ on $Y$. For any shift of $W$, we can isolate the direct effect of $W$ on $Y$ from its indirect effect through the selection. Moreover, we also identify the dependence between $Z$ and $U$ via the function $g$. \\
\indent We observe the joint distribution of $(Y,D, W, Z)$.
Identification relies on the fact that the model and the assumptions imply that, for $u \in [0, 1]$, $w \in \mathcal{W}$, and $z \in \mathcal{Z}$, we have
\begin{equation} \label{main_system}
\begin{aligned}
F_{U|Z}(u|z)&= \quad \P( U \leq u| W=w, Z=z) \\
&= \quad \E[\E[\mathbbm{1}(U \leq u)| D,W, Z]| W=w, Z=z] \\
&= \quad \E[\E[\mathbbm{1}(Y_{DW}\leq h(D,W,u))| D,W, Z]| W=w, Z=z] \\
&= \quad \P(Y\le h(D,w,u) |W=w,Z=z). %\\
\end{aligned}
\end{equation}
The first equality comes from $ U \indep W | Z$ because $V\indep (Z,W)$. Then, we rewrite the equation using the law of total expectation and the third equality follows from the monotonicity of $h$ in its last argument. The last equality is a consequence of the fact that $Y=Y_{DW}$ and of the law of total expectations. %\\

Moreover, for $u \in [0, 1]$, we have
\begin{equation}
\label{added_equation}
\begin{aligned}
u &=\E[F_{U|Z}(u|Z)],\\
%	F_{U|W}(u|w)&= \E[F_{U|Z}(u|Z)|W=w].
\end{aligned}
\end{equation}
by the law of total expectation. This gives one more equation.\footnote{There are more equations of the form $\P(U\le u|W=w)= \E[F_{U|Z}(u|Z)|W=w]$ but they do not help for identification since they introduce as many unknowns as new equations.}

The function $h$ and the conditional distribution functions $F_{ U|Z}$ are identified from the joint distribution of $(Y,D, W, Z)$ if they are the unique solution in a given (possibly infinite-dimensional) parameter space to the continuum of integral equations in \eqref{main_system}-\eqref{added_equation}.
In the next two sub-subsections, we give relevance conditions guaranteeing the uniqueness of the solution to \eqref{main_system}-\eqref{added_equation} when $D$ is discrete (Section \ref{sec.ID_discrete}) and when $D$ is continuous (Section \ref{sec.ID_cont}).

\subsubsection{Identification with discrete $D$}\label{sec.ID_discrete}

Let us assume in this section that $D$ is discrete with $\mathcal{D}=\{1,..., |\mathcal{D}|\}$.
Without loss of generality, we can also suppose that $W$ and $Z$ have discrete support $\mathcal{W}=\{1, ..., |\mathcal{W}|\}$ and $\mathcal{Z} = \{1, ..., |\mathcal{Z}|\}$.
In this setting, \eqref{main_system}-\eqref{added_equation} become

\begin{align}\label{system_discrete}
\begin{split}
F_{U|Z}(u|z) &= \sum_{d=1}^{|\mathcal{D}|} \P(Y\le h(d,w,u), D=d|W=w,Z=z) \\ %\nonumber \\
u &= \sum_{z=1}^{|\mathcal{Z}|}F_{U|Z}(u|z) \P(Z=z ),
\end{split}
\end{align}
for all $u \in [0, 1]$, $w \in \mathcal{W}$, and $z \in \mathcal{Z}$. 
For each $u\in [0,1]$, we obtain a system of $|\mathcal{Z}|\times |\mathcal{W}|+1$ equations, for $|\mathcal{D}| \times |\mathcal{W}| + |\mathcal{Z}|$ unknowns: the $|\mathcal{D}| \times |\mathcal{W}|$ outcome functions' values $h(d,w,u)$ and the $|\mathcal{Z}|$ conditional probabilities $F_{U|Z}(u|z)$. %and the $|\mathcal{W}|$ distributions $F_{U|W}(\cdot|w)$ of $ U | W=w$.
Provided that there are more equations than the number of unknowns, we can proceed to the identification of our objects of interest by solving this system of equations. \\
\indent Let us provide a sufficient identification condition. For integers $N$, $N_1$ and $N_2$, let $\mathcal{I}_N$ be the identify matrix of size $N$ and $\mathcal{O}_{N_1\times N_2}$, the $N_1\times N_2$ matrix whose all entries are equal to $0$. Define  the selection probabilities
$$p_{d|u, w, z} =   \P(D=d |  U = u, W=w, Z=z) .$$
 \begin{comment}
\begin{align*}
	p_{d|u, w, z} = \begin{cases}  \P(D=d |  U = u, W=w, Z=z) \text{ if } f_{U|Z}(u| z) > 0, \\
	0 \text{ otherwise if } f_{U|Z}(u|z) = 0, \end{cases}
\end{align*}
\end{comment}

%$p_{d|u, w, z} = \P(D=d |  U = u, W=w, Z=z)$ the selection probabilities.
\begin{assumption}[Relevance with discrete $D$]\label{ass_relevance} The $$(1+|\mathcal{Z}|\times|\mathcal{W}|)\times (|\mathcal{Z}|+|\mathcal{D}|\times |\mathcal{W}|)$$ matrix of selection probabilities  
\begin{align*}
		\quad M(u)& = \begin{bmatrix}
		% Z = 0
		\mathcal{P}_{W1}(u) & \mathcal{O}_{|\mathcal{Z}|\times |\mathcal{D}|} & \hdots & \mathcal{O}_{|\mathcal{Z}|\times |\mathcal{D}|}  & - \mathcal{I}_{|\mathcal{Z}|}  \\ 
		 \mathcal{O}_{|\mathcal{Z}|\times |\mathcal{D}|}  & \mathcal{P}_{W2}(u) &  \hdots & \vdots &- \mathcal{I}_{|\mathcal{Z}|} \\
		\vdots & \hdots & \ddots &  \mathcal{O}_{|\mathcal{Z}|\times |\mathcal{D}|}  & \vdots  \\
		 \mathcal{O}_{|\mathcal{Z}|\times |\mathcal{D}|}&  \hdots &  \mathcal{O}_{|\mathcal{Z}|\times |\mathcal{D}|} & \mathcal{P}_{W|\mathcal{W}|}(u) & - \mathcal{I}_{|\mathcal{Z}|}  \\ 
				0 & \hdots & \hdots & 0 &\mathcal{P}_Z  \\
	\end{bmatrix}, \\
	\text{where } \quad 
	\mathcal{P}_{Ww}(u) &= \begin{bmatrix}
		p_{1|u, w, 1} & p_{2|u, w, 1}& \hdots & p_{|\mathcal{D}||u, w, 1} \\
		p_{1|u, w, 2} & p_{2|u, w, 2} & \hdots & p_{|\mathcal{D}||u, w, 2} \\
		\vdots & \vdots & \vdots & \vdots \\
		p_{1|u, w, |\mathcal{Z}|} & p_{2|u, w, |\mathcal{Z}|} & \hdots & p_{|\mathcal{D}||u, w, |\mathcal{Z}|}
	\end{bmatrix} \text{ is } |\mathcal{Z}|\times |\mathcal{D}|\\
\text{and}  \quad 
	\mathcal{P}_{Z} &= \begin{bmatrix}
		\P(Z=1) & \P(Z=2) & \hdots & \P(Z=|\mathcal{Z}|) \\
	\end{bmatrix},
\end{align*}
has full column rank for all $u \in [0,1] $. 
\end{assumption}
\noindent This type of full rank conditions are standard when establishing identification of quantile models with endogeneity \citep[see][]{chernozhukovhansen2005, vuongxu2017, feng2024matching}. The condition can be relaxed to allow for a finite set of $u\in[0,1]$ on which the matrix $M(u)$ has a rank-one deficiency \citep[see][]{bruneel2022,bruneel2023don}. Assumption \ref{ass_relevance} is a condition on the underlying selection probabilities conditional on the unobservable $U$. In the case where $|\mathcal{D}| = | \mathcal{W} | = 2$, there are easy-to-interpret sufficient conditions for this assumption to hold. We discuss this below.  \\
%We discuss below easy to interpret sufficient conditions for Assumption \ref{ass_relevance} in the case $|\mathcal{D}| = | \mathcal{W} | = 2$.

\noindent \textbf{Relevance when $|\mathcal{D}| = | \mathcal{W} | = 2$.} The full column rank identification Assumption \ref{ass_relevance} is a relevance condition on the effect of the variables on the true selection probabilities given the unobservable $U$. Let us focus on the simpler case where $D$ and $W$ are binary with $\mathcal{D}=\mathcal{W}=\{1, 2\}$. In this case, a direct necessary condition for Assumption \ref{ass_relevance} is that $|\mathcal{Z}| \geq 3$, because otherwise, the system would have fewer equations than unknowns, so $M(u)$ would not be full column rank. Let us assume $\mathcal{Z}=\{1,2,3\}$. Then, we have $|\mathcal{Z}| \times |\mathcal{W}| + 1 = 7$ equations for $|\mathcal{Z}| +|\mathcal{D}| \times |\mathcal{W}| = 7$ unknowns (four functions, and three conditional distribution functions), and one can show that
\begin{equation}\label{irrelevant_condition}\begin{aligned}
\text{det}\big(M(u)\big) &= p_{1|u, 1, 1} \big( p_{1|u, 2, 2} - p_{1|u, 2, 3} \big) \\
&\quad
+ p_{1|u, 1, 2}\big( p_{1|u, 2, 3}- p_{1|u, 2, 1} \big)
+ p_{1|u, 1, 3} \big( p_{1|u, 2, 1} - p_{1|u, 2, 2}\big).
\end{aligned}
\end{equation}
According to Assumption \ref{ass_relevance}, the model will be identified if $\text{det}\big(M(u)\big) \neq 0$. What does this mean in this case with binary $D$ and $W$? First, $\text{det}\big(M(u)\big) =0$, if $W \indep D | Z, U$, or, $Z \indep D | W, U$. In particular, it means that $Z$ must be relevant for $D$ but not only because of its dependence on $ U$. This is the reason why we refer to the excluded quasi-IV, $Z$, as a "relevant proxy" because it is a proxy for $U$ that is relevant for $D$ even when controlling for $U$. 
In addition, it also requires that $W$ and $Z$ are jointly relevant in the sense that their interaction has an effect on the selection. For example, if $\P(D=1|W=w, Z=z, U=u) = a_u +b_u \mathbbm{1}(w=2) + c_u \mathbbm{1}(z=2) + d_u \mathbbm{1}(z=3)$, there is no joint effect of $W$ and $Z$ on $D$, and one can check that the relevance condition is violated. If we add some interaction terms, e.g., $e_u \mathbbm{1}(w=1, z=2)$, the relevance holds. 
Overall, in this binary treatment case, identification is obtainable if, at every quantile $u$, both quasi-IVs are relevant, and if the effect of a quasi-IV on the selection is heterogenous with respect to the other quasi-IV (joint relevance). \\

The following result establishes (global) identification. 
\begin{theorem}[Identification]\label{identification_theorem}
Suppose that (i) the mappings $y\mapsto \P(Y\le y,D=d|W=w,Z=z)$ are continuously differentiable for all $d\in\mathcal{D}$, $w\in \mathcal{W}$ and $z\in\mathcal{Z}$, (ii) the distribution of $U$ given $Z=z$ is continuous for all $z\in\mathcal{Z}$, (iii) for all $w\in\mathcal{W}$, the distribution of $U$ given $W=w$ has a density $f_{U|W}(u|w)$ such that $ f_{U|W}(u|w)>0$ for all $u\in[0,1]$ and (iv) the mappings  $h(d,w,\cdot),\ d\in\mathcal{D},w\in\mathcal{W}$ are continuously differentiable and strictly increasing. 
Let also Assumption \ref{ass_relevance} hold.
Then,  $h$ and the conditional distribution functions $F_{ U|Z}$ are identified from the joint distribution of $(Y,D, W, Z)$.
\end{theorem} 

\noindent The proof of the theorem can be found in Online Appendix \ref{app_identification_discrete}. Conditions (i) to (iii) in Theorem \ref{identification_theorem} are mild regularity conditions on the joint distribution of $(Y,D,Z,W,U)$. As discussed in Section \ref{section_framework}, the fact that the mappings $h(d,w,\cdot),\ d\in\mathcal{D},w\in\mathcal{W}$ are strictly increasing (condition (iv)) implies rank invariance. The proof of the theorem can be found in Appendix \ref{app_identification_discrete}. The argument goes as follows. First, we show that, under our assumptions, for all $d\in\mathcal{D},w\in\mathcal{W}$, $h(d,w,0)$ is identified as the lower bound of the support of $Y$ given $D=d,W=w$ and, for all $z\in\mathcal{Z}$, $F_{U|Z}(0|z)$ is identified since it is equal to $0$. Then, we demonstrate that there exists a unique solution to the system \eqref{system_discrete} whose value at $u=0$ is given by $h(d,w,0)$ and $F_{U|Z}(0|z)$. The relevance condition ensures that, if we start from a point belonging to the true solution, the optimal path solving the system starting from this point will not deviate from the true solution and will identify the unique solution to the system. \\

% maybe also add $\delta_{wz13}(u)$ ? 

\subsubsection{Identification with continuous $D$}\label{sec.ID_cont}
Let us now consider identification with continuous $D$. We fix $u\in [0,1]$ and introduce $\Epsilon=Y-h(D,W,u)$. Let $f_{\Epsilon|D,Z,W}(e|d,z,w)$ be the density of $\Epsilon$ given $D=d,Z=z,W=w$ at the point $e$ (which existence is guaranteed under our assumptions). We also introduce $\mathcal{F}:\{ F:\mathcal{Z}\to\R:\ \E[F(Z)]=0\}$. For $\Delta\in L^2(D,W)$, we define\footnote{For a random variable $R$, we use the standard notation that $L^2(R)$ is the set of functions $\varphi$ from the support of $R$ to $\R$ such that $\E[\varphi(R)^2]<\infty$.}
$$\omega_\Delta(D,Z,W)=\int_0^1f_{\Epsilon|D,Z,W}(\delta\Delta(D,W)|D,Z,W)d\delta.$$
Let us state the following strong completeness condition and the related identification result.
\begin{assumption}[Relevance with continuous $D$]\label{ass_relevancecont}
For all $\Delta \in L^2(D,W)$ and $f\in\mathcal{F}$,
$$\E\left[\left. \Delta(D,W) \omega_\Delta(D,Z,W)\right| Z,W\right]+f(Z)=0\ \text{a.s.}\Rightarrow \Delta(D,W) =F(Z)=0\text{ a.s..}$$
\end{assumption}

\begin{theorem}\label{ID_contD}Assume that the mapping $(d,w)\in\mathcal{D}\times \mathcal{W}\to h(d,w,u)$ belongs to $L^2(D,W)$ and that the distribution of $\Epsilon$ given $D,Z,W$ is continuous. Let also Assumption \ref{ass_relevancecont} hold. Then, $\Eta\in L^2(D,W)$ and $F\in\mathcal{G}$ solve the system \eqref{main_system}-\eqref{added_equation} if and only if $\Eta(D,W)=h(D,W,u)$ and $F(Z)=F_{U|Z}(u|Z)$ a.s.
\end{theorem}
\noindent The proof is in Section \ref{sec.proof.ID_contID} of the Online Appendix. Assumption \ref{ass_relevancecont} is our model's counterpart of Assumption $L1^*$ in Appendix C of \cite{chernozhukovhansen2005}. Contrary to the previous case with discrete $D$, and as argued in the literature, relevance conditions for nonseparable models with endogenous continuous treatment are difficult to interpret, see the discussions in \cite{canay2013testability} and \cite{beyhum2023instrumental} for the IVQR model. Our condition suffers from the same drawback: it does not have a straightforward meaning.

\section{Additive model}\label{sec.add}

Let us now consider an additive model with homogenous treatment effects. Such a model is interesting because identification results can be obtained in a single framework, allowing for both discrete and continuous $D$. Moreover, the global identification results for continuous $D$ can be derived under more interpretable conditions. We let 
\begin{equation}\label{baseline1add} Y_{dw}=h(d,w)+U,\ \E[U]=0.\end{equation}
This is the counterpart of equation \eqref{baseline1} in mean regression contexts. The residual $U$ does not depend on $d$, meaning that the treatment effects are homogenous (this is the analogue of the rank invariance assumption in the present additive context). We then impose 
\begin{equation}\label{baseline2add} U=g(Z)+V,\ \E[V|Z,W]=0,\end{equation}
which is the analogue of \eqref{baseline2}. Equation \eqref{baseline1add} implies that $\E[U|Z,W]= \E[U|Z]$ so that the conditional independence assumption of Section \ref{sec.ivqr} is now a conditional mean independence condition. 

Here, for $w\in\mathcal{W},z\in\mathcal{Z}$, $h(d,w)+g(z)$ is the average of $Y_d$ given $W=w$ and $Z=z$. As in Section \ref{sec.ivqr}, one could condition on additional covariates $X$. All our results and assumptions can be interpreted as conditional on $X$. 

Let us now turn to identification. We want to identify $h(d,w)$ and $g(z)$ for all $d\in\mathcal{D},w\in\mathcal{W}$ and $z\in\mathcal{Z}$. %To do this, we need a normalization. 
Let us define the (possibily infinite-dimensional) parameter spaces $\mathcal{P}_h$ and $\mathcal{P}_g$ for $h$ and $g$, respectively. As an example, $\mathcal{P}_h$ can be the set of square integrable, bounded or linear functions on the support $\mathcal{S}(D,W)$ of $(D,W)$ to $\R$. 
From the model, we obtain the following system of equations
\begin{equation} \label{Sy_add}\begin{aligned} \E[Y|W,Z]&=\E[h(D,W)+g(Z)|W,Z]\ \text{a.s.},\\
\E[g(Z)]&=0.
\end{aligned}
\end{equation}
The mappings $h$ and $g$ are identified from the joint distribution of $(Y,D,Z,W)$ if this system has a unique solution.\footnote{As in the quantile case, there are additional equations available, of the form $\E[g(Z)|W]=0\ \text{a.s.}$. However, again, such equations are not useful for identification.} 
 This will be the case under the following completeness assumption:
\begin{assumption}[Additive completeness]\label{ass_comp} For all $h\in \mathcal{P}_h$ and $g\in \mathcal{P}_g$,
$$\E[h(D,W)+g(Z)|W,Z]=\E[g(Z)]=0\ \text{a.s.}\ \Rightarrow h(D,W)=g(Z)=0\ \text{a.s.}.$$
\end{assumption}
\noindent Completeness assumptions are standard in the literature on nonparametric instrumental variable models, see \cite{neweypowell2003,darolles2011nonparametric}. In this literature, $D$ is the endogenous variable, $Z$ is the IV and the completeness assumption is different since it corresponds to $
\E[\varphi(D)|Z]=0\ \text{a.s.}\Rightarrow \varphi(D)=0\ \text{a.s.},
$
for all $\varphi\in L^2(D)$. Without the separability of the model, the needed relevance would be $$\E[\varphi(D,W,Z)|W,Z]=0\ \text{a.s.}\ \Rightarrow \varphi(D,W,Z)=0\ \text{a.s.},$$ for functions $\varphi$ in a defined parameter space. This can only hold under degenerate joint distribution of $(D,W,Z)$ or degenerate parameter space. Therefore, the separability of the model allows us to rely on a weaker condition. To further interpret Assumption \ref{ass_comp}, we consider two examples in the following Lemmas. 

\begin{Lemma} \label{lmm_discrete}Suppose $D$, $Z$ and $W$ are discrete with respective supports $\{1,\dots,|\mathcal{D}|\}$, $\{1,\dots,|\mathcal{Z}|\}$, $\{1,\dots,|\mathcal{W}|\}$. For $(d,z,w)\in\mathcal{D}\times \mathcal{Z}\times \mathcal{W}$, let $p_{d|w,z}=\P(D=d|W=w,Z=z)$ and $p_z=\P(Z=z)$. Define the parameter spaces as $\mathcal{P}_h=\left\{h:\mathcal{D}\times \mathcal{W}\to \R\right\}$ and $\mathcal{P}_g=\left\{g:\mathcal{Z}\to \R\right\}$. Assume that the following $(|\mathcal{Z}|\times|\mathcal{W}|+1)\times (|\mathcal{Z}|+ |\mathcal{D}|\times |\mathcal{W}|)$ matrix of selection probabilities
\begin{align*}
		&\quad M= \begin{bmatrix}
		% Z = 0
		\mathcal{P}_{W1} &  \mathcal{O}_{|\mathcal{Z}|\times |\mathcal{D}|}  & \hdots & \mathcal{O}_{|\mathcal{Z}|\times |\mathcal{D}|}  &  \mathcal{I}_{|\mathcal{Z}|} \\ 
		 \mathcal{O}_{|\mathcal{Z}|\times |\mathcal{D}|}  & \mathcal{P}_{W2} &  \hdots & \vdots & \mathcal{I}_{|\mathcal{Z}|} \\
		 \mathcal{O}_{|\mathcal{Z}|\times |\mathcal{D}|}  & \hdots & \ddots & \mathcal{O}_{|\mathcal{Z}|\times |\mathcal{D}|} & \vdots \\
		 \mathcal{O}_{|\mathcal{Z}|\times |\mathcal{D}|}  & \hdots & \mathcal{O}_{|\mathcal{Z}|\times |\mathcal{D}|}  & \mathcal{P}_{W|\mathcal{W}|} & \mathcal{I}_{|\mathcal{Z}|}\\
		0 & \hdots & \hdots & 0 & \mathcal{P}_{Z}
	\end{bmatrix}, \\
	\text{with } \quad 
	&\mathcal{P}_{Ww} = \begin{bmatrix}
		p_{1| w, 1} & p_{2|w, 1}& \hdots & p_{|\mathcal{D}|| w, 1} \\
		p_{1| w, 2} & p_{2| w, 2} & \hdots & p_{|\mathcal{D}|| w, 2} \\
		\vdots & \vdots & \vdots & \vdots \\
		p_{1| w, |\mathcal{Z}|} & p_{2| w, |\mathcal{Z}|} & \hdots & p_{|\mathcal{D}|| w, |\mathcal{Z}|}
	\end{bmatrix}\text{ is }|\mathcal{Z}|\times |\mathcal{D}|,\\
	\text{and}  \quad 
	\mathcal{P}_{Z} &= \begin{bmatrix}
		\P(Z=1) & \P(Z=2) & \hdots & \P(Z=|\mathcal{Z}|) \\
	\end{bmatrix},
	\end{align*}
has full column rank. Then, Assumption \ref{ass_comp} holds. 

\end{Lemma}

\begin{Lemma}\label{lmm_polynomial} Let $h_1,h_2,\dots,h_{r_h}$ and $g_1,g_2,\dots,g_{r_g}$ form families of functions of $L^2(D,W)$ and $L^2(Z)$, respectively. We consider the parameter spaces:
\begin{align*}
\mathcal{P}_h &= \left\{ \sum_{j=1}^{r_h} \beta_j h_j,\ \beta\in\R^{r_h}\right\};\\
\mathcal{P}_g&=\left\{  \sum_{j=1}^{r_g} \alpha_j g_j,\ \alpha\in\R^{r_g}\right\}.
\end{align*}
Suppose that there exists orthonormal functions $\ell_1,\dots,\ell_{r_\ell}$ in $L^2(W,Z)$ such that the $(r_\ell+1)\times (r_h+r_g)$ matrix
\begin{align*}
		&\quad M= \begin{bmatrix}
		% Z = 0
		e^h_{1| 1} & e^h_{2|1}& \hdots & e^h_{r_h| 1}  & e^g_{1|1}  & e^g_{2|1}& \hdots & e^g_{r_g| 1} \\ 
		e^h_{1| 2} & e^h_{2|2}& \hdots & e^h_{r_h| 2}  & e^g_{1|2}  & e^g_{2|2}& \hdots & e^g_{r_g| 2} \\ 
		\vdots & \vdots & \vdots & \vdots & \vdots &\vdots &\vdots &\vdots\\
		e^h_{1| r_\ell} & e^h_{2|r_\ell}& \hdots & e^h_{r_h| r_\ell}  & e^g_{1|r_\ell}  & e^g_{2|r_\ell}& \hdots & e^g_{r_g| r_\ell} \\ 
		0& 0& \hdots & 0 & \E[g_1(Z)]  & \E[g_2(Z)]& \hdots & \E[g_{r_h}(Z)] \\ 
	\end{bmatrix}, \\
		\text{where }
	& \quad e^h_{j| k}=\E[h_j(D,W) \ell_k(W,Z)] \text{ and } e^g_{j| k}=\E[g_j(Z) \ell_k(W,Z)],
\end{align*}
has rank equal to $r_h+r_g$. Then, Assumption \ref{ass_comp} holds. 
\end{Lemma}
\noindent The proof of the lemmas can be found in Section \ref{sec.proof_additive} of the Online Appendix. The rank condition in Lemma \ref{lmm_discrete} is the counterpart of Assumption \ref{ass_relevance} in the additive model. It can be interpreted similarly to Assumption \ref{ass_relevance}. Lemma \ref{lmm_polynomial} allows $D$ to be continuous. It considers parameter spaces consisting of flexible linear combinations of functions. By letting $r_h$ and $r_g$ go to infinity, one can approach nonparametric parameter spaces. An important special case of Lemma \ref{lmm_polynomial} is the linear case, which we elaborate on below.

Let us now formally state our identification theorem. 
\begin{theorem}\label{thm_ID_add} Let Assumption \ref{ass_comp} hold, then 
$\tilde h\in\mathcal{P}_h$ and $\tilde g\in\mathcal{P}_g$ solve \eqref{Sy_add} if and only if $\tilde h(D,W)=h(D,W)$ and $\tilde g(Z)=g(Z)$ almost surely. \\
\end{theorem}
It is proved in Online Appendix \ref{sec.proof_additive}.

\noindent \textbf{Special case: Linear model.} We conclude this section by discussing a special linear case of particular interest. Take $\mathcal{P}_h =\{(d,w)\in\mathcal{D}\times\mathcal{W}\mapsto \alpha+ \beta_Dd +\beta_Ww\}$ and $\mathcal{P}_{g}=\{z\in\mathcal{Z}\mapsto \alpha_Z + \beta_Zz\}$. Then, model \eqref{baseline1add}-\eqref{baseline2add} becomes:
$$Y=(\alpha+\alpha_Z)  + \beta_DD +\beta_WW+  \beta_ZZ +V,\ \E[V|Z,W]=0.$$
The above model is overidentified, and in this special case, $Z$ and $W$ play a perfectly symmetric role. It is known that one could use technical instruments such as $W^2$ or $Z^2$ to identify such a model \citep{bun2019ols,gao2023iv}. \\
\indent The natural counterpart of our nonparametric identification strategy suggests using the interaction $Z \times W$ as an IV to identify this linear model. Importantly and contrarily to the approach using technical instruments, note that the exclusion of $Z \times W$ from the outcome equation is not an additional hard-to-justify assumption \textit{per se}. Indeed, the separability between $Z$ and $W$ is naturally resulting from their properties as valid complementary quasi-IVs. It is a direct implication of the general nonparametric identification arguments in this linear model. In other words, arguing for the complementary validity of $Z$ and $W$ using economic arguments justifies the use of their interaction, $Z\times W$, as a valid IV in the linear model. Providing economic arguments for the quasi-IVs complementarity amounts to providing arguments for their separability. This is particularly useful for practitioners who are generally reluctant to use arbitrary separability assumptions for identification. In a sense, it is the IV counterpart of how the DiD literature justifies the exclusion of interacted fixed effects between time and group assignments from the outcome equation. \\
\indent More precisely, in the notation of Lemma \ref{lmm_polynomial}, we use
$\ell_1(z,w)=1$, $\ell_2(z,w)=z$, $\ell_3(z,w)=w$ and $\ell_4(z,w)=zw$, which yields the moment conditions
%$$\E[\left\{Y-((\alpha+\alpha_Z)+ \beta_DD +\beta_WW+\beta_ZZ)\right\}h_{j}(Z,W)]=0,\ j=1,\dots,4.$$
\begin{equation}\label{eq.linear} \E\left[\left\{Y-((\alpha+\alpha_Z)+ \beta_DD +\beta_WW+\beta_ZZ)\right\}\left(\begin{array}{c} 1\\ Z\\ W\\ ZW
\end{array}\right)\right]=0.\end{equation}
An empirical counterpart of these moment conditions can be estimated through a two-stage least squares (2SLS) estimator using a constant, $D$, $W$ and $Z$ as independent variables and a constant, $Z$, $W$ and $Z\times W$ as instrumental variables. If $Z$ and $W$ are complementary valid quasi-IVs, then $Z\times W$ is excluded from the outcome equation in the linear model. Identification then requires that $Z \times W$ is relevant (i.e., the joint relevance condition) in order for it to be a valid IV. \\
\indent This simple linear model gives explicit intuitions on how our identification strategy works, and it has the advantage of being easily implementable using standard 2SLS routines. We use this approach in our empirical illustration in Section \ref{sec:illustration}.

\section{LATE with invalid IVs}\label{sec.late}

Let us now discuss how we can identify Local Average Treatment Effects (LATE) with invalid IVs. Our LATE framework is neither nested by nor nesting the previous nonseparable models with rank invariance, and the identification arguments differ, so we discuss this in a separate section. 
Relaxing the exclusion and exogeneity restrictions requires again satisfying a (testable) stronger relevance condition than standard IVs. 
\subsection{LATE model }\label{subsection_late}

Throughout this section, let us focus on the binary case, where $D$ has support $\mathcal{D}=\{0,1\}$. Let us denote by $Z$ an excluded quasi-IV with support $\mathcal{Z}\subset \R$, and $W$ a complementary exogenous quasi-IV with support $\mathcal{W}\subset \R$.
The arguments also naturally extend when $W$ and $Z$ contain several variables, but we do not address this case for simplicity of exposition. Again, the model could include additional covariates $X$, but we do not explicitly introduce them and proceed as if every statement was conditional on $X$ to simplify the exposition. \\
\indent In the spirit of \cite{imbensangrist1994}, let $D(w,z)$ be the potential treatment under exogenous quasi-IV equal to $w\in\mathcal{W}$ and excluded quasi-IV equal to $z\in\mathcal{Z}$. Also, denote by $Y_{dw}$ the potential outcome under treatment $d$ and exogenous quasi-IV equal to $w$.
We have the following model for the (continuous or discrete) potential outcomes for $d \in \{0, 1\}$,
\begin{align}\label{model_LATE}
	Y_{dw} = h_{dw} + U_d, \text{ for all $w\in\mathcal{W}$,}
\end{align}
and a selection equation with a latent index structure with additive separability of the shocks: for all $w\in\mathcal{W}$ and $z\in\mathcal{Z}$,
\begin{align}\label{selection_2311}
 &D^*(w,z) = g(w,z) - V, \\
	\quad  \text{and } \quad &D(w,z)=1 \text{ if } D^*(w,z) \geq 0, \ D(w,z)=0 \text{ otherwise.} \nonumber
\end{align}
We observe $(Y,D,W,Z)$, where $Y=Y_{DW}$ and $D=D(W,Z)$. As in the standard LATE framework \citep{imbensangrist1994, vytlacil2002, heckmanvytlacil2005}, ($U_0, U_1, V$) are general unobserved random variables which may be correlated, yielding endogenous selection. $Z$ does not enter directly the outcome equations, but it is (possibly) entering indirectly through $U_0$ and $U_1$. %Similarly, we do not assume that $Z$ is independent from $V$. 
Otherwise, we impose the same assumptions (naturally adjusted for inclusion of $W$ and endogeneity of $Z$) as \cite{heckmanvytlacil2005}, and refer to this paper for more detailed discussions: 
\begin{assumption}\label{ass_latemodel}The following holds:
\begin{itemize}[leftmargin=.6in]
	\item[(i)] \textit{(Independence): ($U_0, U_1, V$) are independent of $W$ conditional on $Z$.}  
	\item[(ii)] \textit{(Rank condition): $g(W, Z)$ is a nondegenerate random variable conditional on $Z$.} 
	\item[(iii)] \textit{The distribution of $V$ is absolutely continuous w.r.t. Lebesgue measure.} 
	\item[(iv)] \textit{(Finite means) The values of $\mathbb{E}|Y_0|$ and $\mathbb{E}|Y_1|$ are finite.} 
	\item[(v)] \textit{$0 < \textrm{Pr}(D=1) < 1$.} 
\end{itemize}
\end{assumption}
\noindent The independence condition (i) is equivalent to the fact that the potential outcomes and treatments $\{Y_{dw},\ D(w,z),\ d\in\mathcal{D},z\in\mathcal{Z},w\in\mathcal{W}\}$ are independent of $W$ conditional on $Z$.
Under independence (i), the separability of $V$ from $g(W, Z)$ in the latent index selection equation \eqref{selection_2311} yields the monotonicity assumption of \cite{imbensangrist1994} for the exogenous quasi-IV, $W$. In other words, for every $z\in\mathcal{Z}$, either $D(w', z) \geq D(w, z)$ for all $(w, w') \in \mathcal{W}\times\mathcal{W}$ or $D(w', z) \leq D(w, z)$ for all $(w, w') \in \mathcal{W}\times\mathcal{W}$. Crucially, without further assumption, the monotonicity only holds for the exogenous quasi-IV, not the endogenous one. Indeed, $Z$ is generally correlated with $V$, so we do not have additive separability of the effect of $Z$ from the shock $V$ on $D^*$. As a consequence, we only express the LATE for changes of $W$, not changes of $Z$, because, without monotonicity with respect to $Z$, the set of compliers is not easily defined. Assumption \ref{ass_latemodel} (ii) is a general relevance condition but we impose stronger and more precise restrictions later. 

A few remarks are in order. Note that if we assumed that $h_{dw}$ does not depend on $w$, then $W$ would become a valid IV. Similarly, $Z$ may be independent of the shocks $(U_0,U_1,V)$; in that case, it is a valid IV. Also, here, $U_d$ does not depend on $w$. Hence, we assume that the effect of $w$ on the potential outcomes is homogenous. In other words, $Y_{dw'}-Y_{dw}$ is not random. Our procedure is, therefore, robust to homogenous direct effects of $W$ on the potential outcomes. Finally, we stress again that we do not assume that $Z$ is exogenous here. \\
\indent Define the propensity score $P(W,Z) = \P(D=1|W,Z)$. 
Since $V$ is independent of $W$ given $Z$, the setup implies that, for any $w' \neq w\in\mathcal{W}$ and $z\in\mathcal{Z}$, if $P(w', z) = P(w, z)$, then $g(w',z)=g(w,z)$ and therefore $D(w', z) = D(w, z)$ almost surely.  This is the key property implied by monotonicity that we will exploit for identification. %\\

\subsection{Treatment effects of interest with quasi-IVs} 
%From here onwards, let us abstract from $X$ to simplify the exposition, and proceed as if every statement was conditional on $X$. 
Let us discuss how the definition of the usual treatment effects of interest is adjusted with quasi-IVs. \\

\noindent \textbf{LATE.} For any $z \in \mathcal{Z}$, conditional on $Z=z$, for any $w'\ne w \in \mathcal{W}$, with $P(w, z) = p < P(w', z) = p'$, define the counterpart of \cite{imbensangrist1994}'s LATE with invalid IVs as
\begin{align}\label{def_late}
	\Delta_{LATE}(w, w'| z) &= \mathbb{E}[ \ Y_{1w'} - Y_{0w'} \ | \ D(w', z)=1,  D(w, z) = 0, Z=z \ ]. 
\end{align}
%{\color{red} Also possible: $\Delta_{LATE}(w, w' | z)$}.
In other words, this LATE is the average treatment effect (at $W=w'$) over the population for which $Z=z$, and that switches from no treatment to treatment when the value of $W$ goes exogenously from $w$ to $w'$.\footnote{One could define the LATE differently, by comparing $Y_{1w'} - Y_{0w}$, $Y_{1w} - Y_{0w}$ or even $Y_{1w} - Y_{0w'}$. We picked $Y_{1w'} - Y_{0w'}$ to connect with the DiD literature where the focus is on the LATE after the treatment, at time $W=w'$. Note that this choice is arbitrary and innocuous since we identify these alternative definitions of the LATE as well using similar arguments.}
As already explained, we can only define a LATE conditional on $Z$, since monotonicity does not hold with respect to $Z$, it is not possible to identify the effect of exogenously moving $Z$ on the subset of the population taking up treatment because of this change. In a sense, $Z$ behaves similarly to standard covariates in the definition of the LATE. However, its (conditional) exclusion from $Y$ allows us to identify the said LATE.\\ % Finally, remark also that if $W$ is in fact excluded, \eqref{def_late} becomes the original LATE of \cite{imbensangrist1994} (conditional on $Z$).

\noindent \textbf{Direct effect of $\boldsymbol{W}$ on $\boldsymbol{Y}$.}  In contrast to the effect that $W$ has on the selection, define the \textit{direct effect} of $W$ on each potential outcome $Y_d$ as
\begin{align*}
\Delta^W_d(w, w') = h_{dw'} - h_{dw} \quad \text{ for } d = 0,1.
\end{align*}
Note that the direct effect of $W$ on $Y$ is also a parameter of interest by itself. Indeed, suppose that $W$ represents a policy change, for example. In that case, we want to be able to know what the \textit{direct effect} of the policy on the outcome is, net of the effect the policy has on the selection into the treatment (\textit{selection effect}, captured by the LATE parameters). \\

\indent To identify the LATE parameters, we exploit exogenous variations of $W$ that shift the treatment probability. The main difference with the LATE with IVs is that $W$ is included in the outcome here, so we also need to isolate the direct effect of $W$ on $Y$. To do so, we exploit the fact that $Z$ is excluded from the outcome. In Section \ref{sec.MTE} of the Online Appendix, we extend our results to identify general LATE and MTE parameters.%\\ %\\

\subsection{Identification of the LATE}\label{subsec.late}
%\subsubsection{Decomposition of the LATE.} 
Let us study identification of $\Delta_{LATE}(w, w'| z)$ using data on $(Y,D,W, Z)$. Consider $w\ne w' \in \mathcal{W}$ and $z \in \mathcal{Z}$, with $P(w, z) = p < P(w', z) = p'$. To identify $\Delta_{LATE}(w, w'| z) $ it is natural to compare the average outcome of the populations such that $\{W=w',Z=z\}$ and $\{W=w,Z=z\}$. We can compute 
\begin{align}\label{eq_late}
	&\mathbb{E}\big[ Y | W=w', Z=z \big] - \mathbb{E}\big[ Y | W=w, Z=z \big] \nonumber \\
	&= \mathbb{E}\big[ Y_{0w'} + (Y_{1w'} - Y_{0w'}) D | W=w', Z=z \big] - \mathbb{E}\big[ Y_{0w} + (Y_{1w} - Y_{0w}) D | W=w, Z=z \big] \nonumber \\ 
	&= \mathbb{E}\big[ Y_{0w'} + (Y_{1w'} - Y_{0w'}) D(w', z) | Z=z\big] - \mathbb{E}\big[ Y_{0w} + (Y_{1w} - Y_{0w}) D(w, z) | Z=z\big] \nonumber \\ 
	&= \mathbb{E}\big[ (Y_{1w'} - Y_{0w'}) ( D(w', z) - D(w, z) ) | Z=z \big] \quad \Big\} = \textit{ effect on compliers} \nonumber \\ %\Delta_{LATE}(s', s) \times (p'-p) \nonumber \\
	& \quad + \underbrace{\mathbb{E}\big[ (Y_{0w'} - Y_{0w}) (1-D(w,z)) | Z=z \big]}_{\textit{effect of $W$ on $Y_0$}}  + \underbrace{\mathbb{E}\big[ (Y_{1w'} - Y_{1w}) D(w, z) | Z=z \big]}_{\textit{effect of $W$ on $Y_1$}},
%	&\nonumber \quad + \underbrace{\mathbb{E}\big[ (Y_{0w'} - Y_{0w}) (1-D(w,z)) | Z=z \big]}_{\textit{effect on never takers}}\\
%	&\quad  + \underbrace{\mathbb{E}\big[ (Y_{1w'} - Y_{1w}) D(w, z) | Z=z \big]}_{\textit{effect on always takers}},
\end{align}
where the second equality comes from the independence of $W$ conditional on $Z$ Assumption \ref{ass_latemodel} (i), and the third one comes from rewriting $Y_{1w} - Y_{0w} = Y_{1w} - Y_{1w'} + Y_{1w'} -Y_{0w'}+Y_{0w'}- Y_{0w}$. %\\
Moreover, by monotonicity/uniformity, $D(w', z) > D(w, z)$ for all subjects if $p' > p$. Hence, the effect on compliers is almost directly the LATE, since:
\begin{align*}
	&\mathbb{E}\big[ (Y_{1w'} - Y_{0w'}) ( D(w', z) - D(w, z) ) | Z=z \big] \\
	&= \mathbb{E}\big[ Y_{1w'} - Y_{0w'} | D(w', z)=1, D(w, z)=0, Z=z] \  \textrm{Pr}(D(w', z)=1, D(w, z)=0) \\ 
	&= \Delta_{LATE}(w, w'|z) (p'-p).
\end{align*} %$\mathbb{E}\big[ (Y_1(s') - Y_0(s)) ( D(s') - D(s) ) \big] = \mathbb{E}\big[ Y_1(s') - Y_0(s) | D(s')=1, D(s)=0 ] \textrm{Pr}(D(s')=1, D(s)=0) = \Delta_{LATE}(s, s') (p'-p)$. 
\indent The main difference with the standard LATE with valid IVs is the presence of the last two terms in Equation (\ref{eq_late}). They represent the direct effects of exogenously changing $W$ from $w$ to $w'$ conditional on $Z=z$ on each potential outcome. These terms are present because $W$ is (possibly) included here, so, contrary to an IV, changing $W$ does not only affect the sorting of individuals (compliers) into treatment (\textit{selection effect}), but also the outcome (\textit{direct effect}). \\
\indent To identify the LATE with quasi-IVs, one needs to identify separately the direct effect of $W$ from its selection effect. Note that 
\begin{align*}
 	&\mathbb{E}\big[ (Y_{1w'} - Y_{1w}) D(w, z) | Z=z\big] = (h_{1w'} -h_{1w}) \ \E[D(w, z) | Z=z] = \Delta_1^W(w, w') \ p, \\
 	\text{ and } & \mathbb{E}\big[ (Y_{0w'} - Y_{0w}) (1-D(w, z)) | Z=z\big] = \Delta_0^W(w, w') (1-p). %+ \Delta_1^Z(z, z') \ p. 
 \end{align*}
 %is indeed directly related to the direct effect of $W$ on $Y_1$ defined previously. 
 So, if we identify $\Delta_0^W(w, w')$ and $\Delta_1^W(w, w')$, we can isolate the two additional terms in \eqref{eq_late}, and obtain the LATE, $\Delta_{LATE}(w, w'|z)$. \\
\indent To identify these direct effects, we use the fact that $Z$ is excluded from $h_{dw}$ in the potential outcome equations. More precisely, we can show that $\Delta_d^W(w, w')$ (for $d=0,1$) are identified under the following local irrelevance assumption. 

\begin{assumption}[Local Irrelevance]\label{ass_localirrelevance}
There exists $z^*\in\mathcal{Z}$ such that, $P(w, z^*) = P(w', z^*)$. 
\end{assumption}

The local irrelevance assumption requires that there exists a $z^*$ such that, switching $W$ from $w$ to $w'$ has no selection effect on the treatment. This holds if $W$ has no effect on $D$ conditional on $Z=z^*$.  In the DiD context, it means that there exists a group for which the same individuals select into treatment before (at $W=w$) and after (at $W=w'$) the treatment occurred \citep[such an assumption is made in][as discussed below]{dechaisemartindhaultfoeuille2018}. More generally, when $W$ has a large support, it only requires that the specific switch from $w$ to $w'$ has no effect on $D$ conditional on $Z=z^*$. In other words, this assumption requires that the effect of $W$ and $Z$ on $D$ is sufficiently nonlinear. This holds if, for example, a change in $W$ has two opposite effects on the treatment, such that for some specific individuals ($Z=z^*$), these effects can perfectly offset each other at some specific values $W=w$ and $W=w'$. 
Such an assumption is more likely to hold when $Z$ is continuous, but it does not require that $W$ or $Z$ are continuous. As mentioned in the introduction, in a different context, \cite{DHAULTFOEUILLE2024105075} also relies on a local irrelevance condition and shows the power of such a restriction. 
We also note that Assumption \ref{ass_localirrelevance} is testable.

Under local irrelevance, Assumption \ref{ass_localirrelevance}, we identify $\Delta_1^W(w, w')$ and $\Delta_0^W(w, w')$. The intuition is that, at $Z=z^*$, under local irrelevance there is no selection effect when we exogenously change $W$ from $w$ to $w'$. As a consequence, the only observable difference in outcome when $W$ changes is only due to the direct effect of $W$ on $Y$ and we identify $\Delta^W_d(w, w')$. 
Formally, at $Z=z^*$ such that $P(w, z^*) = P(w', z^*) = p^*$, we have $D(w, z^*) = D(w', z^*)$ a.s. conditional on $Z=z^*$ by monotonicity. This implies that 
 \begin{align*}&(h_{1w'}-h_{1w})P(w',z^*)= (h_{1w'}-h_{1w})\E[D(w,z^*)|Z=z^*]\\
 &=\E[ (Y_{1w'} - Y_{1w}) D(w, z^*) | Z=z^*]\\
 &=  \E[ Y_{1w'} D(w', z) | Z=z^*]- \E[ Y_{1w'} D(w, z) | Z=z^*]\\
  &= \E[ YD |W=w', Z=z^*\big]- \E\big[ YD | W=w,Z=z^*],
 \end{align*}
So, $\Delta^W_1(w, w') = h_{1w'}-h_{1w}$ is identified by 
$$\Delta^W_1(w, w') =(\E[ YD |W=w', Z=z^*\big]- \E\big[ YD | W=w,Z=z^*])/P(w',z^*),$$
%$$\Delta^W_1(w, w') = (h_{1w'}-h_{1w})=(\E[ YD |W=w', Z=z^*\big]- \E\big[ YD | W=w,Z=z^*])/P(w',z^*),$$
when $P(w',z^*)>0$. Similarly,  $\Delta^W_0(w, w')$ is identified by
$$\Delta^W_0(w, w') =\frac{\E[ Y(1-D) |W=w', Z=z^*\big]- \E\big[ Y(1-D) | W=w,Z=z^*]}{1-P(w',z^*)},$$
when $P(w', z^*) < 1$. We can plug these into \eqref{eq_late} and identify the $\Delta_{LATE}(w, w'|z)$ accordingly. Hence, we have the following theorem. \\

\begin{theorem}[LATE identification] Let Assumptions \ref{ass_latemodel} and \ref{ass_localirrelevance} hold. Suppose that $0<P(w,z^*)<1$, where $z^*$ is defined in Assumption \ref{ass_localirrelevance}. Then, $\Delta_{LATE}(w, w'|z)$ is identified for all $z\in\mathcal{Z}$ such that $P(w',z)>P(w,z)$. \\
\end{theorem}

\noindent \textbf{Comparison with fuzzy DiD.} In the case where $W\in\{0,1\}$ is time (with two time periods) and $Z\in\{0,1\}$ is the group assignment (with two groups), it is helpful to compare our results with that of \cite{dechaisemartindhaultfoeuille2018} for fuzzy DiD.  This paper makes two crucial sets of assumptions. First, there are the common trends and conditional common trends conditions (Assumptions 2 and 4' in this paper). These assumptions are both implied by our model \eqref{model_LATE}. Formally, our model \eqref{model_LATE} is more restrictive, but in the special case of fuzzy DiD, we could relax it to just impose some conditions on the trends as \cite{dechaisemartindhaultfoeuille2018}. The advantage is that we allow for more general variables than group and time, which are potentially continuous.
Second, they assume that the treatment rate does not change in the control group between the two periods (Assumption 2 in their paper). This is our local irrelevance Assumption \ref{ass_localirrelevance} in the special case where $Z=z^*$ is the control group and $w=0$, $w'=1$ are the dates. As previously discussed, with continuous $W$ the local irrelevance is more general than this. \\ %It also allows to identify more general LATE and MTE parameters, as we discuss in Section \ref{sec.MTE}.  %\\ 

\noindent \textbf{Marginal Treatment Effects and Generalized LATE.} Following the same intuition, Appendix \ref{sec.MTE} proves how to identify MTE and more general LATE parameters with quasi-IVs.

\section{Empirical illustration}\label{sec:illustration}

To illustrate how to use quasi-IVs in practice, we estimate household Engel curves. An Engel curve studies the share of total expenditure of the household that is spent on a given class of goods: food expenditures here (\citealp{banks1997quadratic, blundell2007semi, horowitz2011, deaner2023proxy,beyhum2023one}, for a short survey see \citealp{lewbel2008engel}). \\
\indent Unfortunately, one cannot estimate the Engel curve by regressing the share of food consumption on the total expenditures because of endogeneity concerns. Indeed, consumption is a choice, and typically, the unobservable household preference for food should impact both the total consumption and the share of food consumed at a given level of total consumption.
A solution that has been employed to address this endogeneity is to use household income as an IV \citep{blundell2007semi}. 
However, one may be concerned that income is also endogenous with respect to unobservable preferences. Indeed, everything else equal, a household with a stronger preference for food consumption may decide to work more to increase its food consumption (even at a fixed level of total expenditures). In a fully general structural model, the household labor decisions (and thus their income) should depend on their preferences for every class of goods. As a consequence, income is possibly an invalid IV. \\

\noindent \textbf{Choice of quasi-IVs.} 
To address endogeneity, we rely on two complementary quasi-IVs instead. As an excluded quasi-IV, $Z_{it}$, we use the (lagged) household income. As discussed above, it is possibly endogenous because it may be correlated with the unobserved preferences for food.\footnote{To be more precise, denote $U_{it}$ the unobserved preference for food, and $I_{it}$ the household income at time $t$. It is likely that $I_{it}$ is determined as a function of $U_{it}$ (because households adjust how much they work accordingly). Moreover, the preferences for food are likely correlated over time, so $U_{it}$ is correlated with $U_{it-1}$, and as a consequence, $I_{it-1}$ is also endogenous with respect to $U_{it}$.} On the other hand, conditional on these unobservables and on the total expenditures, it should have no direct effect on the share of food consumption, so it is credibly excluded. Moreover, the (lagged) income should also be relevant to the household's total expenditures. \\
\indent We complement this excluded quasi-IV, with exogenous unexpected shocks (in percentage changes) to the local market (at the state level) earnings as an exogenous quasi-IV, $W_{it}$. These local market shocks are credibly independent of the unobservable household preference for consumption as well as from the lagged household earnings.\footnote{This is the reason why we use the lagged earnings and not the current earnings: otherwise, the current earnings may not be independent of these shocks. While with the lagged earnings, we credibly have that $W_{it} \indep (U_{it}, Z_{it})$.} These local shocks are also relevant because they directly impact the current total expenditures through their effects on the current earnings. However, we cannot \textit{a priori} rule out that they have a direct impact on the share of food consumed. Indeed, similarly to how consumption responds differently to permanent or transitory earnings shocks(\citealp{blundell2008consumption, commault2022does}, for a survey see \citealp{crawley2024income}) households may choose to spend unexpected earnings shocks disproportionately on some type of goods (e.g., leisure or food in restaurants). As a consequence, unexpected earnings shocks may directly affect the share of food consumed, not only due to their pass-through to the overall consumption, so these are possibly invalid IVs.\footnote{One can also use other local shocks, such as shocks to the prices of specific goods, or simply local inflation level as exogenous shocks here. } \\

\noindent \textbf{Data.} We use data from the $1999$ to $2014$ waves of the Consumer Expenditure Surveys (CEX) from the U.S. Bureau of Labor Statistics to construct our household (real) consumption and (real) income data.\footnote{Note that this dataset is not a panel per se. However, there are several interviews. To construct the "lagged" income, we use the household income reported in the first interview. At the same time, all the other variables (consumption) are taken from the last interview several months later, i.e., at the end of the year.}\footnote{We also ran a similar analysis on the Panel Study of Income Dynamics (PSID) and obtained very similar results, but with fewer observations per year. } To construct the unexpected shocks to earnings, we use the panel of per capita personal income by state from the Bureau of Economic Analysis (BEA).\footnote{To construct "unexpected" shocks, we proceed as follows. Denote $I_{st}$ the average nominal income in state $s$ at time $t$. We construct observed shocks $\Delta I_{st} = (I_{st} - I_{st-1})/I_{st-1}$.
%\begin{align*}
%	\Delta I_{st} = \frac{I_{st} - I_{st-1}}{I_{st-1}}.
%\end{align*}
Then, we run the two-way fixed effects regression to obtain expected shocks, $\widehat{\Delta I_{st}} = \delta_t + \delta_s$, and use the residuals as the unpredicted earnings shocks, i.e., $W_{it} = \Delta I_{st} - \widehat{\Delta I_{st}}$.
%\begin{align*}
%	\text{Earnings Shocks}, W_{it} = \Delta I_{st} - \widehat{\Delta I_{st}}.
%\end{align*}
}
We focus on a homogenous subsample of households composed of a couple with two dependent children. We do not control for any other covariates (except for time and state fixed effects). We also remove outliers with yearly (real) consumption and (real) income above $\$300,000$ per year or consumption below $\$10,000$ per year (less than $2\%$ of the overall sample). We end up with a total of $3,948$ observations. \\

 \noindent \textbf{Estimation.} For simplicity, we estimate a simple linear relationship between the (log of) total expenditures and the share spent on food (inside the household and outside). We use a 2SLS estimation procedure to solve the moments given by equation \eqref{eq.linear}. This allows us to illustrate the simplicity of the use of quasi-IVs in a linear model. Indeed, as already shown in Section \ref{sec.add}, in the linear case, the interaction $Z \times W$ serves as an IV for $D$, and we include both $Z$ and $W$ linearly in the model for the outcome (and for the treatment). As a consequence, in this specific case, the estimation can be carried out via a standard 2SLS regression routine, which is especially nice for practical implementation. Again, as stressed out previously, the linear model can be estimated differently (using technical instruments for e.g.), but using $Z\times W$ as an IV is the direct implication of our general nonparametric identification results to this specific linear model. Our general nonparametric identification results provide a rationale for the use of the interaction as an instrument. \\
\indent Note also that in this linear model, exogenous and excluded quasi-IVs are undistinguishable because they both appear in a similar manner in the outcome equation. This may be a confusing feature of the linear model and is not generally the case in our other more general models. Our identification arguments obviously do not rely on this. Here, our quasi-IVs narrative provides a rationale for the separability of the effect of $Z$ and $W$ on $Y$, and thus, a rationale for why $Z \times W$ should not appear in the outcome equation (and can thus be a valid IV, provided that joint relevance is satisfied). \\
\begin{table}[h!] \centering 
\caption{Engels Curve estimation}\label{tab:engels}
\scalebox{0.9}{
  \begin{tabular}{@{\extracolsep{5pt}}lccccc} 
\\[-1.8ex]\hline 
\hline \\[-1.8ex] 
%& \multicolumn{4}{c}{\textit{Dependent variable:}} \\ 
\\[-1.8ex] & \multicolumn{5}{c}{\textbf{Outcome}, $Y_{it}$: Share of food consumption} \\ 
%\cline{2-4} \cline{5-7} 
\\[-1.8ex] & \textbf{OLS} & \multicolumn{2}{c}{\textbf{IV}} & \multicolumn{2}{c}{\textbf{quasi-IVs}}    \\ 
\cline{2-2} \cline{3-4} \cline{5-6}
\\[-1.8ex] &  & $1$st stage & $2$nd stage  & $1$st stage & $2$nd stage     \\ 
\hline \\[-1.8ex] 
%  & & & & & \\ 
    & & & &  \\  
\textbf{Treatment}, $D_{it}$:  &  &  & &  \\
log(Total consumption$_{it}$)  & -0.0692$^{***}$ & & -0.0721$^{***}$  &  & -0.1702$^{***}$\\
 & (0.0021) &  & (0.0036) &  & (0.0344)   \\ 
    & & & &  \\

\textbf{Excluded quasi-IV}, $Z_{it}$ & &  & & & \\
log(Income$_{it-1}$) & & 0.3901$^{***}$ & & 0.3939$^{***}$ & 0.0383$^{**}$ \\ 
 &  & (0.0084) &  & (0.0084) & (0.0135)   \\ 
& & & & & \\

\textbf{Exogenous quasi-IV}, $W_{it}$ & &  & & &  \\
Local earnings shocks$_{it}$ & & & & -30.43$^{***}$& -0.0077 \\ 
& & & & (5.2127) & (0.0812) \\
& & & & & \\

%\textbf{Instrumental Variable} & &  & & \\
%lag(income) & & & & \\ 
%& & & & \\

\textbf{Interaction $W_{it} \times Z_{it}$}  & &  & & 2.7197$^{***}$ & \\
(for quasi-IVs' $1$st stage) & & & & (0.4747) &  \\
& & & & & \\
\hline \\[-1.8ex] 
Number of observations & \multicolumn{5}{c}{3,948}  \\ 
%Weak Instrument F-stat & & & $2168.226^{***}$ & & $32.82^{***}$ \\ 
\hline 
\hline \\[-1.8ex] 
\multicolumn{6}{l}{\textit{Note:} $^{*}$p$<$0.1; $^{**}$p$<$0.05; $^{***}$p$<$0.01. } \\
\multicolumn{6}{l}{In every regressions we control for state and year fixed effects. } \\
\multicolumn{6}{l}{We use real income and consumption (base $2000$). } \\
%\multicolumn{6}{l}{In the IV regression, we use the (lag) income as the IV. } \\ 
\end{tabular}
}
\end{table} 
%
%\indent The results of the estimation with quasi-IVs are provided in Table \ref{tab:engels}. We also report the results one obtains by running the (naive) OLS estimator on the regression of the share of food consumption on the (log) total expenditures and the corresponding IV regression where one uses our excluded quasi-IV (the lagged income) as a valid IV.

\noindent \textbf{Results.} 
\noindent As is standard in this literature, in all specifications, we find that the share of food consumed decreases with the total expenditures: as households become richer, they consume less food (in proportion). We also find that the baseline ordinary least squares (OLS) estimate of the effect of consumption on the share of food consumed is less negative than the IV estimate.
Our quasi-IVs estimate is much more negative than both the OLS and  IV estimates. For an increase of $1\%$ total consumption, we estimate with quasi-IVs that the share of food decreases by $0.1702$ percentage points. The result is significant at the 1\% level. This is to be compared with the estimated decreases of $0.0692$ and $0.0721$ percentage points obtained with the OLS and IV estimation procedures, respectively. The fact that we find decreasing effects more than twice as large as the ones estimated by OLS suggests an important endogeneity problem that households with higher preference for food consumption also consume more overall. We also find a difference with the IV estimates because it seems that the lagged income is not a valid IV. Indeed, as visible with the results of the quasi-IV regression, the lagged income has a direct significant effect on the share of food consumed, suggesting that households with a higher preference for food also choose to work more (and earn more) in order to increase their food consumption. \\
\indent While the estimates clearly indicate that the lagged income is not excluded, and thus not a valid IV, we find that the exogenous local earnings shocks have no significant direct effect on the share of food consumed. They only affect it through their effect on the total consumption. It was not possible to know this \textit{a priori}, but our quasi-IV estimation suggests (or at least, it does not reject) that these exogenous $W$ could be used as valid IVs as well. This illustrates how the validity of quasi-IVs can provide an alternative way to test the validity (exclusion) of IVs \citep[compared to known methods as in][for example]{kitagawa2015}. \\
\indent The "relevance cost" of the estimation with quasi-IVs also appears clearly in this illustration. As explained before, in the linear model, it is as if the estimation procedure uses the interaction $W\times Z$ as a standard IV for $D$. As a consequence, we need this interaction to have a relevant effect on the treatment (the total consumption here). Fortunately, this is a testable restriction, and we find that the interaction is relevant here. However, it also appears clearly that it is not always obvious that the interaction is going to be relevant, and in general, it is often less relevant than the two variables $Z$ and $W$ taken separately. This may result in a problem of weak IV and exploding standard errors in the second stage. The instruments do not seem to be weak here (we reject that the interaction is a weak IV, with a large F-stat of $32.82$), but the standard error of the coefficient of the treatment is around 10 times larger than the one of the corresponding IV estimate. One solution to circumvent this issue is to use more interactions as IVs by including additional $W$ shocks (shocks to local prices, for example).

\section{Conclusion} \label{sec.ccl}

In this paper, we have shown that quantities typically identified with an IV can be identified using two complementary quasi-IVs instead. Several economic examples
of quasi-IVs satisfying our assumptions have been given, demonstrating that our approach opens new identification strategies for applications. In the simple linear case,
the estimation of treatment effects with quasi-IVs can be carried out by running a standard 2SLS using the interaction between the complementary quasi-IVs as a valid
IV. However, for the more general models presented in the paper, estimation is a natural and crucial next step before bringing the methodology to the data. Our identification proofs are constructive and, in this sense, can be used to define an estimator. Nevertheless, several computational challenges will arise from the complex nature of the problem at hand. We will tackle these estimation challenges in future research.\\
\indent Another avenue is to study identification with other types of invalid IVs that complement each other in a different manner than the exogenous and excluded quasi-IVs studied here. The semi-IVs of \cite{bruneel2023don} are an example, but there may be other empirically relevant cases.

\section*{Acknowledgments}
We are especially grateful to Geert Dhaene for his extensive comments. We also thank Laurens Cherchye, Frank Kleibergen, Xavier d'Haultfoeuille, Alexandre Gaillard, Yagan Hazard, Art$\bar{\text{u}}$ras Juodis, Thierry Magnac, Martin Mugnier, Lorenzo Navarini, Florian Schuett, Gabriela Miyazato Szini, Frank Verboven, Frederic Vermeulen, Lina Zhang and seminar participants at the \textit{Universiteit van Amsterdam} for useful comments. We acknowledge financial support from the KU Leuven Research Funds through the grants STG/21/040 and STG/23/014.

\setlength{\bibsep}{0pt}

\bibliographystyle{apalike}

%\footnotesize
%\setlength{\baselineskip}{0pt} 
\bibliography{references}

%\appendix
%\noindent {\LARGE \textbf{Supplementary materials: online appendices}} %\\

\end{document}

%\newgeometry{top=0.8in} 

\title{{\fontsize{14}{20} \selectfont 
%\vspace{1cm}
\textbf{%Don't (fully) exclude me, it's not necessary! \\
Online Appendix to ``Identification with possibly invalid IVs''}}}

\author{{\fontsize{12}{20} Christophe Bruneel-Zupanc}\footnote{E-mail address: \href{mailto:christophe.bruneel@gmail.com}{christophe.bruneel@gmail.com}} \\ {\small \textit{Department of Economics, KU Leuven}} \and {\fontsize{12}{20} Jad Beyhum}\footnote{E-mail address: \href{mailto:jad.beyhum@kuleuven.be}{jad.beyhum@kuleuven.be}}\\ {\small \textit{Department of Economics, KU Leuven}} } 
%\vspace{0.25cm} %grantSTG/21/040.} \\

%\vspace{10cm}

\date{{\normalsize \today}\vspace{0.25cm} \\
%{\color{blue}{\normalsize \centering \textbf{Preliminary and Incomplete!}  \textbf{Please do not circulate}}}
%\parbox{\linewidth}{\centering \textbf{Preliminary and Incomplete!}  \textbf{Do not circulate}}
}

%\vspace{-10cm}
\maketitle

%\vspace{-0.20in}

\onehalfspacing % check that gives 32 lines/page. Seems ok. 

{\hypersetup{linkbordercolor=black,linkcolor = black,
			urlcolor  = blue,
			filecolor=black}
		% or \hypersetup{linkcolor=black}, if the colorlinks=true option of hyperref is used
		\tableofcontents
	}

\newpage

\section{General LATE and MTE}\label{sec.MTE}
In this section, we show how we can identify more general treatment parameters in this model using intuitions similar to the ones used in Section \ref{subsec.late}. We work with the model and the assumption of Section \ref{sec.late}. Let us define the parameters of interest.\\
\subsection{Parameters of interest}
\noindent \textbf{Generalized LATE.} Let us normalize $V|Z=z \sim \mathcal{U}(0, 1)$ for any $z \in \mathcal{Z}$.\footnote{This normalization is innocuous given the assumptions of the model. Indeed, suppose the latent variable generating the choices is $D^* = \nu(W, Z) - \Eta$, where $\Eta$ is a general continuous random variable. In that case, one can always reparametrize the model such that $g(W, Z) = F_{\Eta|Z}(\nu(W, Z))$ and $V=F_{\Eta|Z}(\Eta)$. } This does not mean that $V$ is independent from $Z$, it is just a normalization ensuring that $P(w,z)=g(w,z)$ for all $w\in\mathcal{W},z\in\mathcal{Z}$.\footnote{This is because, under this normalization, $P(w,z)=\P(D=1|W=w,Z=z)= \P(V\le g(w,z)|W=w,Z=z)= \P(V\le g(w,z)|Z=z)=g(w,z)$.} Everything only holds conditional on $Z$, and it only allows to compare $g(W, Z)$ for variations of $W$ given $Z$, but not to compare $g(W, Z)$ across different realizations of $Z$. In this sense, $Z$ really acts as standard covariates $X$ in the LATE literature \citep[e.g.,][]{heckmanvytlacil2007b}, where everything holds conditional on $X$, the normalization of $V$ is conditional on $X$, and we cannot compare across $X$. The only difference is that contrary to covariates $X$, $Z$ is excluded from $h_{dw}$, allowing us to identify $h_{dw}$. \\
\indent Under this normalization, following \cite{heckmanvytlacil2005}, we can define a more general LATE, for any $w \in \mathcal{W}, z \in \mathcal{Z},\ p\le p' \in [0,1]^2$, as
\begin{align*}
	\Delta_{LATE}(w, p, p'|z) = \mathbb{E}\Big[ Y_{1w} - Y_{0w} \big| V \in [p, p'], Z=z \Big]. 
\end{align*}
This definition is more general and, as its counterpart with IVs, allows the distinction between the definition of the parameter and its identification. First, if $P(w', z)=p' > P(w, z)=p$, $\Delta_{LATE}(w', p, p'|z) = \Delta_{LATE}(w, w'|z)$ since the events $\{D(w', z)=1,\ D(w, z)=0\}$ and $\{P(w, z) < V \leq P(w', z)\}$ are equivalent. Hence, the general LATE can be interpreted as the previous one. More generally, $\Delta_{LATE}(w, p, p'|z)$ represents the average treatment effect on individuals, with $W=w$ and $Z=z$, who would be induced to switch into treatment if the treatment probability exogenously changed from $p$ to $p'$. This LATE can be considered a specific policy-relevant treatment effect (PRTE).\\

\noindent \textbf{Marginal Treatment Effect (MTE).} We can naturally define the MTE \citep[][]{heckmanvytlacil2005} as the limit case of the general LATE when $p' \rightarrow p$, i.e., 
\begin{align*}
	\Delta_{MTE}(w, p |z) &= \mathbb{E}\Big[ Y_{1w} - Y_{0w} \big| V = p, Z=z \Big] \\
	&= \underset{p'\rightarrow p}{\text{lim}}\ \Delta_{LATE}(w, p, p'|z).
\end{align*}
The MTE gives the expected effect of treatment for individuals with $W=w, Z=z$ who would be indifferent between $D=1$ and $D=0$ if $V=p$. Again, the MTE is defined independently of its identification, and we will show that it can be identified even if $P(w, z) \neq p$. \\

\subsection{Identification}
To obtain identification, we first use the local irrelevance condition to identify the direct effect of the exogenous quasi-IV on the outcome. Once it is identified, we can identify the LATE and MTE given $(w, z)$ at specific probabilities $p$, because we can disentangle the selection effect of exogenously changing $W$, from its direct effect on the outcome. \\

\noindent \textbf{Identification of the direct effect of the exogenous quasi-IV.} %\\
To identify $\Delta^W_d(w, w')$ for $d=0,1$, we rely on the observable propensity score $P(w,z)$ to check the local irrelevance condition at a given $z\in\mathcal{Z}$.
First, if there exists a $Z=z^*$ such that $P(w, z^*) = P(w', z^*)$, then $\Delta^W_d(w, w')$ is identified for both $d=0,1$, as described earlier. But this does not stop here because we can also identify $\Delta^W_d(w, w')$ indirectly. For example, if there exist $z^*_1\in\mathcal{Z}$ such that $0\le P(w_0, z^*_1) = P(w_1, z^*_1)\le 1$ and another $z^*_2\in\mathcal{Z}$ such that $0\le P(w_1, z^*_2) = P(w_2, z^*_2)\le 1$, then $\Delta^W_d(w_0, w_2)$ is identified indirectly, as the sum of $\Delta^W_d(w_0, w_1)$ and $\Delta^W_d(w_1, w_2)$ which are directly identified. This idea can be generalized to multiple indirect connections, yielding the following Theorem.

\begin{theorem}[Identification direct effect of $W$]
Consider $w,w'\in\mathcal{W}$ such that there exists a sequence of elements of $\mathcal{Z}$, denoted $\{z^*_k\}_{k=1,...,K-1}$ and of intermediary elements of $\mathcal{W}$, denoted $\{w_k\}_{k=1,...,K}$ with $w_1 = w$ and $w_K = w'$, such that $0\le P(w_k, z_k) = P(w_{k+1}, z_k)\le 1$. Then, $\Delta^W_d(w, w')$ is identified for $d\in\{0,1\}$.
\end{theorem}

\indent The existence of these sequences of locally irrelevant points is directly visible in the data, from the identified propensity score $P(w,z),\ w\in \mathcal{W},\ z\in\mathcal{Z}$. Denote the set of identified $\Delta_d^W(w, w')$ starting from $w$ as
\begin{align*}
I_{W}(w) = \{ w' \in \mathcal{W}: \Delta_d^W(w, w'),\ d\in\mathcal{D}, \text{ is identified} \}.
\end{align*}
This set is directly identified from the observable data on the propensity score $P(w,z),\ w\in \mathcal{W},\ z\in\mathcal{Z}$.
In general, if there exists a $z^*$ and a $\tilde{w}$ such that $P(w, z^*) = P(\tilde{w}, z^*)$, if $P(w,z)$ is smooth in $w$ and $z$, $W$ and $Z$ are continuous random variables, and if there is sufficient nonlinearity in the effect of $W$ on $D$ with respect to $Z$, $I_W(w)$ should be a large subset of $\mathcal{W}$. \\

\noindent \textbf{Identification of the general LATE.} Using the identified direct effect of $W$ on $Y$, we can identify general LATE parameters. %\\

\begin{theorem}[General LATE identification] For any $w \in \mathcal{W}, z \in \mathcal{Z}$, and $p, p' \in [0, 1]^2$, if there exists $\tilde{w}$ and $\tilde{w}' \in I_W(w)$ with $P(\tilde{w}, z) = p$ and $P(\tilde{w}', z) = p'$, then $\Delta_{LATE}(w, p, p'|z)$ is identified. %\\
\end{theorem}

\noindent \textbf{Proof.} Under the assumptions of the theorem, we have 
\begin{align*}
	&\mathbb{E}[Y|W=\tilde{w}', Z=z] - \mathbb{E}[Y | W=\tilde{w}, Z=z] \\
	&=\mathbb{E}[Y_{0\tilde{w}'} + (Y_{1\tilde{w}'} - Y_{0\tilde{w}'}) D(\tilde{w}', z) | Z=z] - \mathbb{E}[Y_{0\tilde{w}} + (Y_{1\tilde{w}} - Y_{0\tilde{w}}) D(\tilde{w}, z) | Z=z] \\
	&= \ \Delta_{LATE}(w, p, p') (p' - p) \\
	& \quad + \Delta^W_1(w, \tilde{w}') p' \ - \Delta^W_1(w, \tilde{w}) p + \Delta^W_0(w, \tilde{w}') (1-p') \ - \Delta^W_0(w, \tilde{w}) (1-p),
\end{align*}
where the first equality comes from the definition of $Y$ and Assumption \ref{ass_latemodel} (i) and the second equality follows from rewriting $Y_{d\tilde{w}} = Y_{d\tilde{w}} - Y_{dw} + Y_{dw}$ and $Y_{d\tilde{w}'} = Y_{d\tilde{w}'} - Y_{dw} + Y_{dw}$. As a consequence, $\Delta_{LATE}(w, p, p')$ is identified as a function of observables and previously identified objects by
\begin{align*}
	&\Delta_{LATE}(w, p, p')  \\
	& = \frac{\mathbb{E}[Y|W=\tilde{w}', Z=z] - \mathbb{E}[Y | W=\tilde{w}, Z=z]}{p'-p} \\
	& \quad + \frac{-\Delta^W_1(w, \tilde{w}') p' \ + \Delta^W_1(w, \tilde{w}) p - \Delta^W_0(w, \tilde{w}') (1-p') \ + \Delta^W_0(w, \tilde{w}) (1-p)}{p'-p}. \quad \quad \quad  \blacksquare
\end{align*}
\indent Note that $\Delta_{LATE}(w, p, p'|z)$ can be identified even if $P(w, z) \neq p$ and $P(w, z) \neq p'$. It can even be identified if there exists no $\tilde{z}$ such that $P(w, \tilde{z}) = p$ or $p'$. Indeed, because we identify the effect of $W$ on $Y$, we can isolate the LATE, which represents the pure effect of switching the selection probability from $p$ to $p'$ at a specific value $w$ of $W$. \\

\noindent \textbf{Identification of the MTE.} Taking the previous arguments to the limit, we also identify the MTE as the limit case of the general LATE.

\begin{theorem}[MTE identification] For any $w \in \mathcal{W}, z \in \mathcal{Z}$, and $p \in [0, 1]$, if there exists $\tilde{w}$ in the interior of $I_W(w)$ with $P(\tilde{w}, z) = p$ and $\partial P(\tilde{w}, z)/\partial w \neq 0$, then $\Delta_{MTE}(w, p|z)$ is identified.
\end{theorem}

\indent Note that the identification of the MTE requires that $W$ and $Z$ are continuous and the propensity score is smooth to ensure that $I_W(w)$ is continuous. If $I_W(w)$ is large, the set of identified MTE should also be large. As for the general LATE, identifying $\Delta_{MTE}(w, p|z)$ does not require that $P(w, z) = p$, or even that there exists a $\tilde{z}$ such that $P(w, \tilde{z}) = p$. The MTE is often not discussed in the DiD literature because it requires continuous $W$ and $Z$. With quasi-IVs, it is not a problem since many examples of quasi-IVs are continuous.

\section{Potential outcomes justification of model \eqref{baseline1}-\eqref{baseline2}}\label{sec.alternative}
In the baseline framework given by \eqref{baseline1}-\eqref{baseline2} in the main text, $U$ can be seen as a structural error term that has a clear structural interpretation (e.g., unobserved ability, preference, productivity, ...).
Alternatively, we could write the problem in terms of potential outcomes. Let $F_{Y_d|W}(\cdot|w)$ be the cumulative distribution function of $Y_d$ given $W=w$. Assume that $F_{Y_d|W}(\cdot|w)$ is strictly increasing. Then, we can write
%Our framework can also be interpreted in terms of potential outcomes with
\begin{align}
\label{baseline1b} Y_d&=\tilde{h}(d,W,\tilde{U}_d),\ \tilde{U}_d | W \sim\mathcal{U}[0,1],
\end{align}
where $h(d,w,u)= (F_{Y|d|W}(\cdot|w))^{-1}(u)$ and $\tilde{U}_d = F_{Y_d|W}(Y_d|W)$ is the rank of $Y_d$ in its distribution given $W$. By construction, $\tilde{U}_d \indep W$. Then, we impose rank invariance, which means that there exists a variable $\tilde{U}$ such that $\tilde{U}_d = \tilde{U}$ for all $d\in\mathcal{D}$. This variable $ \tilde{U}$ may depend on $Z$, and we assume that
$$ \tilde{U}\indep W|Z,$$
which allows us to write
% Remark: more in Draft_3.
\begin{align}
\label{baseline2b} \tilde{U}&=\tilde{g}(Z, \tilde{V}),\ \tilde{V} | Z,W \sim\mathcal{U}[0,1],
\end{align}
where $\tilde{g}(z,v)=\inf\{u\in[0,1]:\ F_{U|Z}(u|z)\ge v\}$ for all $z\in\mathcal{Z},v\in[0,1]$, $\tilde{V}= F_{\tilde{U}|Z}(\tilde{U}|Z)$ and $F_{\tilde{U}|Z}(\cdot|z)$ is the cumulative distribution function of $\tilde{U}$ given $Z=z$.

Very importantly, notice that $\tilde{U} \indep W $, which is guaranteed by construction, is nonnested with the assumption $\tilde{U} \indep W | Z$. 
The baseline framework \eqref{baseline1}-\eqref{baseline2} actually fits into this framework written in terms of potential outcomes and would yield $\tilde{U} \indep W | Z$ (even though $U$ is generally different from $\tilde{U}$ since we do not impose $U\indep W$). The two frameworks can be analyzed through the same identification arguments.
%This framework written in terms of potential outcomes, is ultimately equivalent to our main Framework when it comes to the identification arguments. 
\section{Proof of Theorem \ref{identification_theorem}}\label{app_identification_discrete}
We introduce further notation.  Let $\mathcal{Y}$ be the support of $Y$ and $\mathcal{H}=\mathcal{Y}^{|\mathcal{D}|\times|\mathcal{W}|} \times [0,1]^{ |\mathcal{Z}|}$. Let us also define $h_{dw}(u) = h(d, w, u)$ and the vector of all the functions of interest:
\begin{align*}
	\mathbf{h}(u) = \Big[ h_{11}(u) \cdots h_{|\mathcal{D}|1}(u) \cdots h_{1|\mathcal{W}|}(u) \cdots h_{|\mathcal{D}||\mathcal{W}|}(u) \quad F_{U|Z}(u|1) \cdots F_{U|Z}(u| |\mathcal{Z}|) \Big].
\end{align*}
Then, $\mathbf{h}(u)$ solves the system of equations $G(\boldsymbol{\Eta}, u)=0$ for all $u\in[0,1]$ where $G:\mathcal{H}\times [0,1]\mapsto \R^{|\mathcal{W}|\times |\mathcal{Z}|+1}$ and 
$$ \left(G(\boldsymbol{\Eta},u)\right)_{wz}=\sum_{d=1}^{|\mathcal{D}|}\P(Y\le \Eta_{d w}, D=d|W=w, Z=z)-\Eta_{z}^F,$$
for all $d\in\mathcal{D}$ and $w\in \mathcal{W}$ and 
$$ \left(G(\boldsymbol{\Eta},u)\right)_{|\mathcal{W}|\times |\mathcal{Z}|+1}=\sum_{z=1}^{|\mathcal{Z}|}\Eta_{z}^F \P(Z=z)-u,$$
with 
$$\boldsymbol{\Eta}=\Big[\Eta_{11}(u)  \cdots \Eta_{|\mathcal{D}|1}(u) \cdots \Eta_{1|\mathcal{W}|}(u) \cdots \Eta_{|\mathcal{D}||\mathcal{W}|}(u) \quad \Eta^F_1\cdots \Eta^F_{|\mathcal{Z}|}\Big]\in\mathcal{H}.$$
For $d\in\mathcal{D},w\in\mathcal{W},z\in\mathcal{Z}$, let also $f_{Y,d|w, z}(y)$ and $f_{U,d|w,z}(u)$ be, respectively, the derivative of the mapping $y\in\R\mapsto \P(Y\le y,D=d|W=w,Z=z)$ at the point $y$ and the derivative of the mapping $u\in\R\mapsto \P(U\le u,D=d|W=w,Z=z)
$ at the point $u$. \\

\noindent \textit{Step 1.}
We first show that $\mathbf{h}(0)$ is identified. First, notice that we know that
$$F_{U|Z}(0|1)=\dots= F_{U|Z}(0| |\mathcal{Z}|)=0$$ since we assumed that the distribution of $U$ given $Z$ is continuous. Moreover, the fact that $M(0)$ has full column rank implies that $\P(D=d,W=w|U=0)>0$ for all $d\in\mathcal{D}$ and $w\in \mathcal{W}$. For all $d\in\mathcal{D},w\in\mathcal{W}$, since $h(d,w,\cdot)$ is increasing, this yields that $h(d, w, 0)$ is identified as the lower bound of the support of $Y$ given $D=d,W=w$. This shows that $\mathbf{h}(0)$ is identified.\\

\noindent \textit{Step 2.} Next, we show that the Jacobian of the mapping $ G(\cdot,u)$ at the point $\mathbf{h}(u)$ exists and has full column rank for all $u\in[0,1]$. Since the mappings $y\mapsto \P(Y\le y,D=d|W=w,Z=z),\ d\in\mathcal{D},w\in\mathcal{W},z\in\mathcal{Z}$ are differentiable, this Jacobian is given by the $\left( |\mathcal{Z}|\times |\mathcal{W}|+1\right)\times \left(|\mathcal{D}|\times |\mathcal{W}|+|\mathcal{Z}|\right)$ matrix

\begin{align*}
		&\quad \nabla_{\boldsymbol{\Eta}} G(\mathbf{f}(u),u) = \begin{bmatrix}
		% Z = 0
		\mathcal{F}_{W1}(\mathbf{h}(u)) &  \mathcal{O}_{|\mathcal{Z}|\times |\mathcal{D}|}  & \hdots &  \mathcal{O}_{|\mathcal{Z}|\times |\mathcal{D}|}  & - \mathcal{I}_{|\mathcal{Z}|}  \\ 
		 \mathcal{O}_{|\mathcal{Z}|\times |\mathcal{D}|}  & \mathcal{F}_{W2}(\mathbf{h}(u)) &  \hdots & \vdots &- \mathcal{I}_{|\mathcal{Z}|} \\
		\vdots & \hdots & \ddots &  \mathcal{O}_{|\mathcal{Z}|\times |\mathcal{D}|}  & \vdots  \\
		 \mathcal{O}_{|\mathcal{Z}|\times |\mathcal{D}|}  & \hdots & \hdots & \mathcal{F}_{W|\mathcal{W}|}(\mathbf{h}(u)) & - \mathcal{I}_{|\mathcal{Z}|}  \\ 
				0 & \hdots & \hdots & 0 &\mathcal{P}_Z  \\
	\end{bmatrix}, \\
	&\text{where }
	\mathcal{F}_{Ww}(\mathbf{h}(u)) = \begin{bmatrix}
		f_{Y,1 | w, 1}(h_{1w}(u))  & f_{Y,2 | w, 1}(h_{2w}(u))& \hdots & f_{Y,|\mathcal{D}| | w, 1}(h_{|\mathcal{D}|w}(u)) \\
		f_{Y,1 | w, 2}(h_{1w}(u))  & f_{Y,2 | w, 2}(h_{2w}(u))& \hdots & f_{Y,|\mathcal{D}| | w, 2}(h_{|\mathcal{D}|w}(u)) \\
		\vdots & \vdots & \vdots & \vdots \\
		f_{Y,1 | w, |\mathcal{Z}|}(h_{1w}(u))  & f_{Y,2 | w, |\mathcal{Z}|}(h_{2,w}(u))& \hdots & f_{Y,|\mathcal{D}| | w, |\mathcal{Z}|}(h_{|\mathcal{D}|w}(u)) \\
	\end{bmatrix}  \\
&\text{ is } |\mathcal{Z}|\times |\mathcal{D}| \text{ and } 
	\mathcal{P}_{Z} = \begin{bmatrix}
		\P(Z=1) & \P(Z=2) & \hdots & \P(Z=|\mathcal{Z}|)
	\end{bmatrix}.
\end{align*}
Next, we have 
$$\P(Y\le y,D=d | W=w,Z=z)=\P(U\le  h_{dw}^{-1}(y),D=d|W=w,Z=z), $$
so that, by the chain rule, it holds that 
\begin{align*}
	%f_{Y,d|w, z}(y)=  f_{U,D|W,Z}(h_{d,w}^{-1}(y)) \ (h_{d,w}^{-1})'(y),
	f_{Y,d|w, z}(y)=  f_{U,d|w,z}(h_{d,w}^{-1}(y)) \ (h_{d,w}^{-1})'(y).
\end{align*}
Bayes' rule and the fact that $U\indep W|Z$ imply that  $f_{U,d|w,z}( u)=\P(D=d|U= u,W=w,Z=z) f_{U|W}( u | w)$. This yields $$f_{Y,d|w, z}(h_{d,w}(u))= p_{d|u,w,z} \  f_{U|W}(  u|w) \ (h_{dw}^{-1})'(h_{dw}(u)).$$ By the inverse function theorem, we have 
$$ (h_{dw}^{-1})'(h_{dw}(u))=\frac{1}{h'_{dw}(u)}.$$
This implies that  $$f_{Y,d|w, z}(h_{dw}(u))= p_{d|u,w,z}  \frac{f_{U|W}(  u|w)}{h'_{dw}(u)}.$$
As a consequence, we can rewrite $ \nabla_{\boldsymbol{\Eta}} G(\mathbf{h}(u),u)$ as 
\begin{align*}
	 \nabla_{\boldsymbol{\Eta}} G(\mathbf{h}(u),u)= M(u)  H\big(\mathbf{h}(u)\big), 
\end{align*}
with $M(u)$ defined in Assumption \ref{ass_relevance} and where $H(\mathbf{h}(u))$ is the diagonal matrix with elements equal to
\begin{align*}
	\left( \frac{f_{U|W}(  u|1)}{h'_{11}(u)},\cdots \frac{f_{U|W}(  u|1)}{h'_{|\mathcal{D}|1}(u)},\cdots,\frac{f_{U|W}(  u||\mathcal{W}|)}{h'_{1|\mathcal{W}|}(u)}\cdots \frac{f_{U|W}(  u||\mathcal{W}|)}{h'_{|\mathcal{D}||\mathcal{W}|}(u)}\quad \underbrace{1 \cdots 1}_{\text{size }|\mathcal{Z}|}\right).
\end{align*}
All the diagonal elements of $H(\mathbf{h}(u))$ are strictly positive (notably because $h_{dw}$ are strictly increasing in $u$). As a consequence, $ \nabla_{\boldsymbol{\Eta}} G(\mathbf{h}(u),u)$ is full column rank since $M(u)$ is full column rank by Assumption \ref{ass_relevance}. \\

\noindent \textit{Step 3.} As a last step we show uniqueness. Let us consider $\tilde{\mathbf{\eta}}(\cdot)$ such that $G(\tilde{\mathbf{\eta}}(u),u)=0,\ u\in[0,1]$ and $\tilde{\mathbf{\eta}}(0)=\mathbf{h}(0)$. Let $\bar u=\sup \{u\in[0,1]:\ \text{ for all } u'\le u,\  \tilde{\mathbf{\eta}}(u')= \mathbf{h}(u')\}$. Let us show that $\bar u=1$. We reason by contradiction. Suppose that $\bar u< 1$. Then, by continuity of the mapping $\nabla_{\boldsymbol{\Eta}}G(\cdot,\cdot)$ at the point $(\mathbf{h}(\bar u),\bar u)$, the continuity of the determinant of a matrix and the fact that $\nabla_{\boldsymbol{\Eta}}G(\mathbf{h}(\bar u),\bar u)$ has full column rank (see previous step), there exists an open ball $\mathcal{B}\subset \mathcal{H}\times [0,1]$ around $(\mathbf{h}(\bar u),\bar u)$ such that, for all $(\boldsymbol{\Eta},\upsilon)$ in $\mathcal{B}$, $\nabla_{\boldsymbol{\Eta}}G(\boldsymbol{\Eta},\upsilon)$ has full column rank. By definition of $\bar u$ and the continuity of $\tilde{\mathbf{\eta}}(\cdot)$ and $\mathbf{h}(\cdot)$ there exists $\tilde u\in[0,1]$ such that $(\tilde{\mathbf{\eta}}(\tilde u),\tilde u)$ and $(\mathbf{h}(\tilde u),\tilde u)$ belong to $\mathcal{B}$ and $\tilde{\mathbf{\eta}}(\tilde u)\ne \mathbf{h}(\tilde u)$.
Applying the mean-value theorem, to $ \nabla_{\boldsymbol{\Eta}}G(\cdot,\tilde u)$ we obtain that there exists $\boldsymbol{\Eta}^*\in\mathcal{H}$ such that $(\boldsymbol{\Eta}^*,\tilde u)\in\mathcal{B}$ and 
$$0=G(\tilde{\mathbf{\eta}}(\tilde u),\tilde u)- G( \mathbf{h}(\tilde u),\tilde u)=\nabla_{\boldsymbol{\Eta}}G(\boldsymbol{\Eta}^*,\tilde u)(\tilde{\mathbf{\eta}}(\tilde u)- \mathbf{h}(\tilde u))\ne 0,  $$ where we used the fact that $\nabla_{_{\boldsymbol{\Eta}}}G(\boldsymbol{\Eta}^*,\tilde u)$ has full column rank by definition of $\mathcal{B}$ and $\tilde{\mathbf{\eta}}(\tilde u)- \mathbf{h}(\tilde u)\ne0$ by definition of $\tilde u$. 
This yields the contradiction. \\

\indent This proof shares similarities with the proof of \cite{feng2024matching} in a different context. Note that the theorem can alternatively be proven by showing that there is a unique solution starting from known $\mathbf{h}(0)$ to the system of differential equation obtained by deriving the original system with respect to $u$. Such an approach is described for similar problems in the IVQR setup in \cite{bruneel2022, bruneel2023don}. \\

\section{Proof of Theorem \ref{ID_contD}}\label{sec.proof.ID_contID}

Let $(\tilde h,\tilde{F}_{Z|U})$ be another admissible solution to the system \eqref{main_system}-\eqref{added_equation}. The following holds
\begin{equation}\label{Syst_cont}
\begin{aligned}
 &\E[\E[\mathbbm{1}(Y \leq h(D,W,u))-\mathbbm{1}(Y \leq \tilde h(D,W,u))|  D,W, Z]| W,Z]\\
 &\quad +F_{Z|U}(u|Z)-\tilde{F}_{Z|U}(u|Z)=0\ \text{a.s.}.\\
&\E[F_{Z|U}(u|Z)-\tilde{F}_{Z|U}(u|Z)]=0\ \text{a.s.},
\end{aligned}
\end{equation}
Let $\Delta_h(D,W) =\tilde h(D,W,u)-h(D,W,u)$ and $\Delta_F(Z)=F_{Z|U}(u|Z)-\tilde{F}_{Z|U}(u|Z)$. By definition of $\Epsilon$, the first equation of \eqref{Syst_cont} is equivalent to
\begin{align*}
&\E[\E[I(\Epsilon\le 0)-\E[I(\Epsilon \leq \Delta_h(D,W))|  D,W, Z]| Z,W]+\Delta_F(Z)=0\ \text{a.s.}\\
&\iff \E\left[\left.\int_0^1 \Delta_h(D,W) f_{\Epsilon|D,Z,W}(\delta \Delta_h(D,W)|D,Z,W)\right| W,Z\right]+\Delta_F(Z)=0\ \text{a.s.}
\end{align*}
The system \eqref{Syst_cont} is then equivalent to 
\begin{equation}
\label{syst_fin}
\begin{aligned}
\E\left[\left. \Delta_h(D,W) \omega_{\Delta_h}(D,Z,W)\right| W,Z\right]+\Delta_F(Z)&=0\ \text{a.s.};\\
\E[\Delta_F(Z)]&=0,
\end{aligned}
\end{equation}
which implies $\Delta_h(D,W)=\Delta_F(Z)=0\ \text{a.s.}$ by Assumption \ref{ass_relevancecont}.

\section{Proof of the results of Section \ref{sec.add}}\label{sec.proof_additive}

\subsection{Proof of Lemma \ref{lmm_discrete}} Take  $\tilde h\in \mathcal{P}_h$ and $\tilde g\in \mathcal{P}_g$ such that
\begin{equation}\label{sy_lmm_dis} \E[\tilde h(D,W)+\tilde g(Z)|W,Z]=\E[\tilde g(Z)]=0\ \text{a.s..}\end{equation}
Let
$
v=[\tilde h(1,1),\dots,\tilde h(|\mathcal{D}|,1),\dots,\tilde h(1,|\mathcal{W}|),\dots, \tilde h(|\mathcal{D}|,|\mathcal{W}|),\tilde g(1),\dots,\tilde g(|\mathcal{Z}|)]^\top.
$
The system of equations \eqref{sy_lmm_dis} can be rewritten as $ M v=0$.
The full column rank assumption on $M$ implies $v=0$.

\subsection{Proof of Lemma \ref{lmm_polynomial}} 
Take  $\tilde h=\sum_{j=1}^{r_h}\beta_j h_j\in \mathcal{P}_h$ and $\tilde g=\sum_{j=1}^{r_j}\alpha_j g_j\in \mathcal{P}_g$ such that
$$ \E[\tilde h(D,W)+\tilde g(Z)|W,Z]=\E[\tilde g(Z)]=0\ \text{a.s..}$$
By the law of total expectations, this implies
\begin{equation}\label{sy_lmm_pol} \E[(\tilde h(D,W)+\tilde g(Z))\ell_j(W,Z)]=\E[\tilde g(Z)]=0\ \text{a.s.},\ j=1,\dots,r_\ell.\end{equation}
Let $v=[\beta_1,\dots,\beta_{r_h},\alpha_1,\dots,\alpha_{r_g}]^\top.$ The system of equations \eqref{sy_lmm_pol} can be rewritten as $ M v=0$. This yields $v=0$ by the full column rank condition on $M$.
\subsection{Proof of Theorem \ref{thm_ID_add}}
Suppose that $\tilde h\in \mathcal{P}_h$ and $\tilde g\in \mathcal{P}_g$ satisfy 
\begin{align*}\E[Y|W,Z]&=\E[\tilde h(D,W)+\tilde g(Z)|W,Z]\ \text{a.s.;}\\
\E[\tilde g(Z)]&=0.
\end{align*}
Then, we obtain 
$$\E[\Delta_h(D,W)|W,Z]+\Delta_g(Z)=\E[\Delta_g(Z)]=0\ \text{a.s.},$$ where $\Delta_h= \tilde h-h$ and $\Delta_g= \tilde g -g$. This yields the result by Assumption \ref{ass_comp}.

\setlength{\bibsep}{0pt}

\bibliographystyle{apalike}

\bibliography{references}

%\appendix
%\noindent {\LARGE \textbf{Supplementary materials: online appendices}} %\\